AAO NEWSLETTER

ANGLO-AUSTRALIAN OBSERVATORY

# The first on-sky demonstration of photonic OH suppression

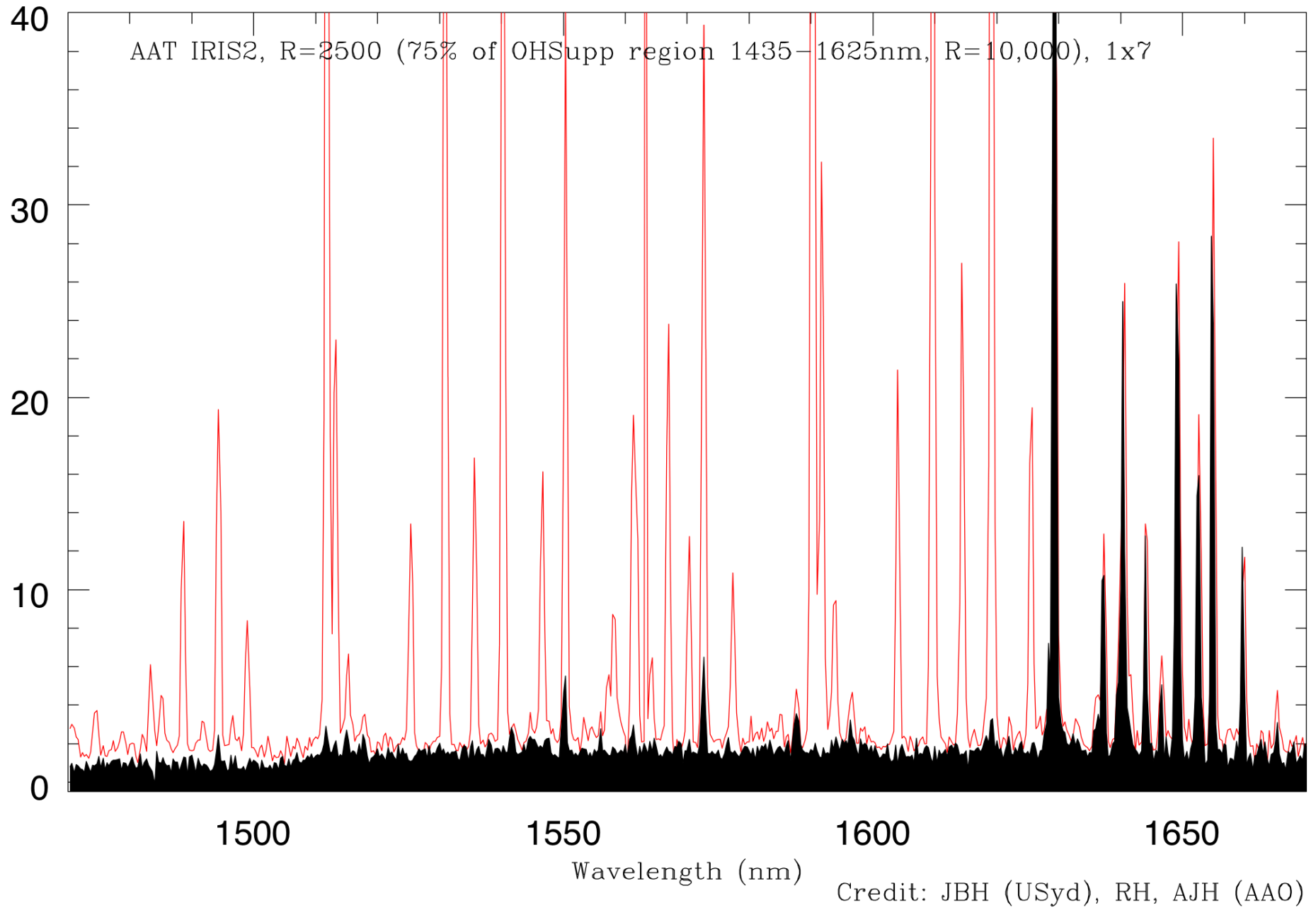

A demonstration of near-IR spectroscopy with OH-suppressing fibres of the night sky, obtained with the AAT and IRIS2 in December 2008. The OH suppressed spectrum is shown in black and the unsuppressed spectrum in red.

contents

# DIRECTOR'S MESSAGE

At the time the last AAO Newsletter went to press, the question of the scientific lifetime of the AAT was being reviewed by an eminent international committee (ANSOC). In their report, they concluded that: "The ANSOC are convinced that there is a world-leading programme of survey astronomy to be carried out by the AAT over the next ten years, until late into the next decade, and that funding should be continued at roughly its present level through that period. The proposals contained in the document *The Anglo-Australian Telescope – Case to ANSOC* are fully endorsed by the Committee and recommended for full funding for a period of ten years through the routes outlined in that document". The AAO's case and the ANSOC report can be found under *What's new?* on the AAO web pages at http://www.aao.gov.au. The review's strong endorsement of the AAO's plans for the future has been immensely encouraging for the Observatory as it works on evolving from a joint Anglo-Australian venture towards becoming Australia's national optical observatory.

A significant part of the AAO's strength is its long-term development of new technologies for astronomical instrumentation. The front page of this Newsletter shows the impressive results from the first on-sky demonstration of OH suppression at near-infrared wavelengths using photonic fibres. These tests, which used IRIS2 on the AAT, are described in the article on page 15 of this Newsletter. Because the OH sky lines are removed at high resolution with negligible leakage and no scattering within the spectrograph, this technique is capable of reducing the observed background between 0.9 and 1.8 microns by a factor of at least 30. The scientific potential of this technology is immense – potential science applications include rest-frame ultraviolet observations of the Universe at the time of first light and reionization. The AAO and collaborators hope to bring this technology to the point of application on large telescopes as early as 2010.

Some of the AAO's other instrumentation programs have also recently passed milestones. The next major instrument being built for the AAT is a high-resolution spectrograph called HERMES, which will be fed by the 400 fibres of the 2dF system. HERMES will be a versatile general-purpose facility enabling survey-style astronomy at high spectral resolutions. The primary science drivers are massive 'Galactic Archaeology' surveys that will unravel the formation history of the Milky Way by measuring kinematics and chemical abundances for very large samples of stars. HERMES went through a conceptual down-select process in September 2008 and will progress to preliminary design review in June this year.

The AAO leads an international team of six institutions that has just completed one of two competing design studies for WFMOS, a wide-field multi-object spectrograph proposed as a joint Gemini/Subaru facility. The WFMOS conceptual design study has just been submitted and the team is currently preparing for a review at the end of February. As proposed, WFMOS would be an extremely powerful instrument, with thousands of fibres accessing a 1.5-degree field of view at the prime focus of the Subaru 8-metre telescope: 3000 fibres would feed ten low-resolution spectrographs, while another 1500 fibres would feed four high-resolution spectrographs. The primary science drivers for WFMOS at low spectral resolution are massive redshift surveys of galaxies at $z\sim1$ and $z\sim3$ that would measure the evolution in the equation of state of the universe, and determine the nature of the dark energy that is causing the acceleration in the Hubble expansion. At high spectral resolution the main science drivers are Galactic Archaeology surveys of not only the Milky Way but also other nearby galaxies such as M31 (Andromeda). However, these extraordinary capabilities are expensive, and it remains to be seen whether Gemini and Subaru will be able to fund such an ambitious project in these times of financial crisis.

As well as developing powerful new instrumentation, the AAO also fosters its special strength in survey astronomy by supporting ambitious large observing programs. Another *Call for Proposals* for new Large Programs starting in semester 09B on the AAT has just been released (see the 'Observing' tab on the AAO web pages). These proposals may use any of the AAT instruments to compellingly address major scientific questions with allocations of telescope time ranging from fifty nights to hundreds of nights over several semesters. At least 25% of the time on the AAT is given over to Large Programs – in fact the AAO encourages ambitious proposals and does not set an upper limit on the fraction of time awarded to Large Programs.

The two Large Programs currently running on the AAT are both producing some early science results. In the last Newsletter (#114, p.3), Driver et al. described the goals and strategy of the Galaxy And Mass Assembly (GAMA) survey. In this Newsletter, the article by Drinkwater et al. on page 3 describes the sample of starburst galaxies being observed for the WiggleZ dark energy survey and shows some of the preliminary results. The primary goal of the survey is to measure the baryon acoustic oscillation 'standard ruler' at redshifts $0.5<z<1.0$ and so obtain the first constraints on the dark energy equation of state at that epoch. The AAO is now looking for the next generation of Large Programs – do you have an exciting large project addressing some important question in mind?

Matthew Colless

# THE WIGGLEZ DARK ENERGY SURVEY: 100,000 STARBURST GALAXY REDSHIFTS

Michael Drinkwater (Queensland) and Chris Blake (Swinburne) for the WiggleZ Team[1]

## Introduction

The WiggleZ Dark Energy Survey is an Australian-led large project on the AAT, which will measure some 200,000 spectroscopic redshifts of distant emission-line galaxies with the AAOmega spectrograph. The aim of the project is to determine the scale of baryon acoustic oscillations (BAO) imprinted on the spatial distribution of the galaxies. The BAO scale is a standard ruler that can be used to measure cosmic distances. Our BAO determination will be accurate to 2 per cent and will constrain theories of dark energy. In this article we describe the design and initial results of the project, which started in Semester 2006B and is scheduled to finish observations in 2010A.

The WiggleZ survey traces its origins to the extraordinary precision with which the key cosmological parameters can now be measured. The age, expansion rate, geometry, matter and energy content of the universe are now determined to a precision of better than 10%; we have entered the era of "precision cosmology" [1]. Paradoxically, this success has revealed an enormous gap in our knowledge of the underlying physics. This was already noted in the early 1990s as it became clear that the Universe was geometrically flat, but the matter density was well below the critical value, requiring an additional (large) contribution from a non-zero cosmological constant term [2]. The need for this additional term was confirmed when studies of distant supernovae revealed that the expansion rate of the universe is accelerating [3,4]. The problem is that we need new physics to explain this acceleration: either gravity is fundamentally different from the vision put forward by Einstein, or the cosmic energy budget is dominated by a new form of matter with a negative pressure – "dark energy" [5]. Several large, high-redshift, galaxy surveys have been proposed to test models of dark energy. The first of these to commence is the WiggleZ survey. In the rest of this article we describe the design of the survey, our current progress and our first results.

## The survey design and current status

We describe the two main features of our survey design in this section. Firstly, our approach is based on a geometrical measurement to make an independent test of dark energy. Secondly, we use ultraviolet satellite data to select high-redshift galaxies with strong emission lines so we can measure their redshifts very rapidly with the AAOmega spectrograph.

A powerful and independent way to measure the effects of dark energy is to use geometrical relations between distance and redshift, as these are a function of the cosmological parameters, notably the nature of the dark energy contribution. For such tests we require a standard ruler, a known physical scale whose observed angular size can be measured as a function of redshift.

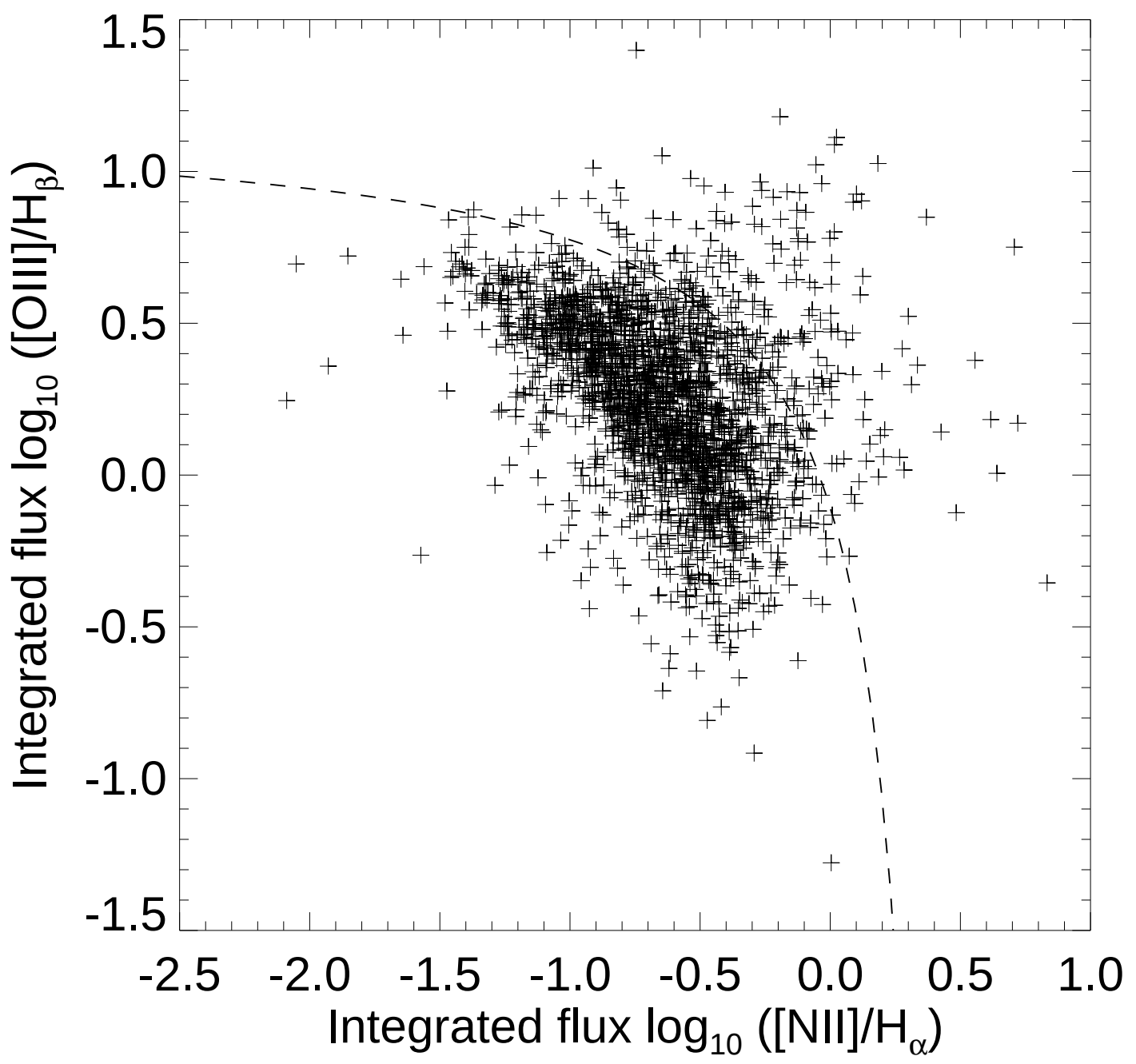

Figure 1: Diagnostic plot of the emission-line ratios of [NII] to Hα and [OIII] to Hβ for WiggleZ galaxies. The dashed line illustrates the canonical division between star-forming galaxies (bottom left of the plot) and AGN (top right of the plot). The WiggleZ galaxies are mainly star-forming.

[1]The WiggleZ team members are: C.A. Blake, S. Brough, W.J. Couch, K. Glazebrook, G.B. Poole, M. Pracy (Swinburne University of Technology); M. Colless, R. Sharp (AAO); S. Croom (University of Sydney); M.J. Drinkwater, T.M. Davis, R.J. Jurek, K.A. Pimbblet (University of Queensland); K. Foster, D.C. Martin, T. Small, T. Wyder (Caltech); B. Madore (Carnegie); D. Woods (University of British Columbia & UNSW).

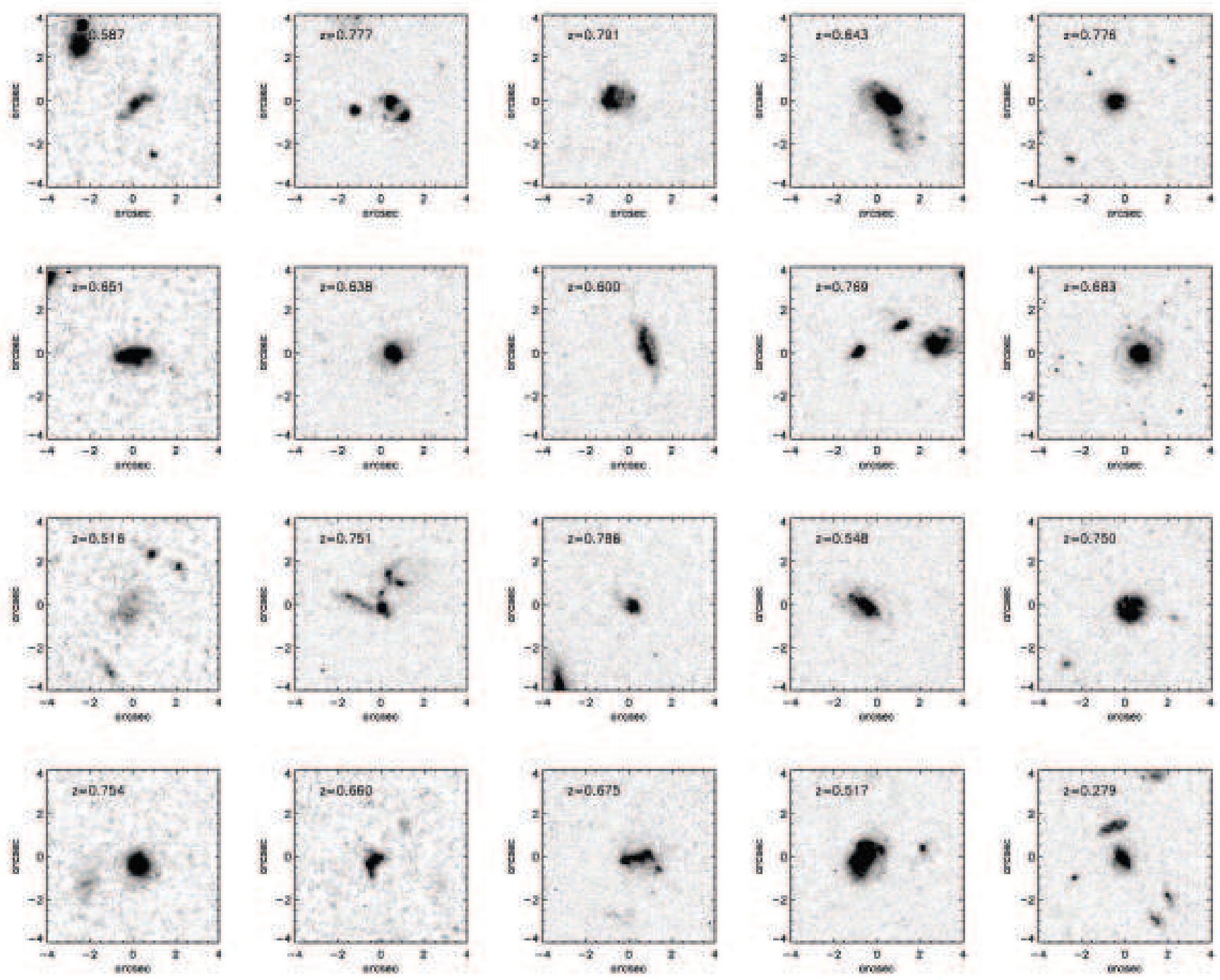

Figure 2: Postage stamp images of WiggleZ targets observed serendipitously by the Hubble Space Telescope. They display a high degree of merger activity.

Fortunately, such a standard ruler exists: the baryon acoustic oscillation (BAO) scale imprinted on the distribution of baryonic matter at recombination in the early Universe. The BAO signal has been measured in the cosmic microwave background [6] (i.e. at recombination) and, more recently, in the present-day galaxy distribution [7, 8]. These galaxy samples are too close to measure dark energy, but they establish the feasibility of the BAO method. Measurement of the BAO scale in the galaxy distribution at high redshifts will give a powerful test of the dark energy models [9].

The WiggleZ target galaxies are selected using ultra-violet photometry from the Galaxy Evolution Explorer (GALEX) satellite [10] in two bands, the *FUV* from 135–175 nm and *NUV* from 175–275 nm. We detect galaxies in the *NUV* filter and then use the *NUV/FUV* ratio of their fluxes to select high-redshift galaxies. This works because the observed *FUV* flux of a galaxy drops rapidly as the redshift increases past $z$=0.5 as the Lyman break in the (rest wavelength) galaxy spectrum shifts into that filter. We use additional optical photometry to further improve the fraction of galaxies with high redshifts. We then observe the targets with AAOmega to confirm that they are galaxies and measure their redshifts. This is a massive undertaking for a survey of this size and it would not be possible without the extremely high efficiency of the AAOmega/2dF system on the AAT. We typically obtain some 2500 galaxy spectra per clear night when observing for WiggleZ.

At the time of writing (2009 January) we have observed 130 scheduled nights, just over half the allocated 220 nights for this long-term survey. We have measured redshifts for some 130 000 galaxies. We will not be able to measure the BAO scale until this survey is complete, but we have already obtained several important results [11] which we summarise in the following sections.

**The nature of the WiggleZ galaxies**

The key result in terms of the feasibility of the project is that we can efficiently measure large numbers of redshifts of the selected galaxies with the AAOmega spectrograph in relatively short 1-hour exposures. The galaxies are extremely faint (optical *R* magnitude around 22), so the strong emission line features in the spectra are essential for the redshift measurements.

The ratios of the strengths of these emission lines can be used to distinguish whether they originate from star formation or from an active galactic nucleus. A diagnostic plot of the line ratios is shown in Figure 1 and demonstrates that star formation is mainly responsible. The emission-line fluxes indicate that the star formation rate in the galaxies is high; typically

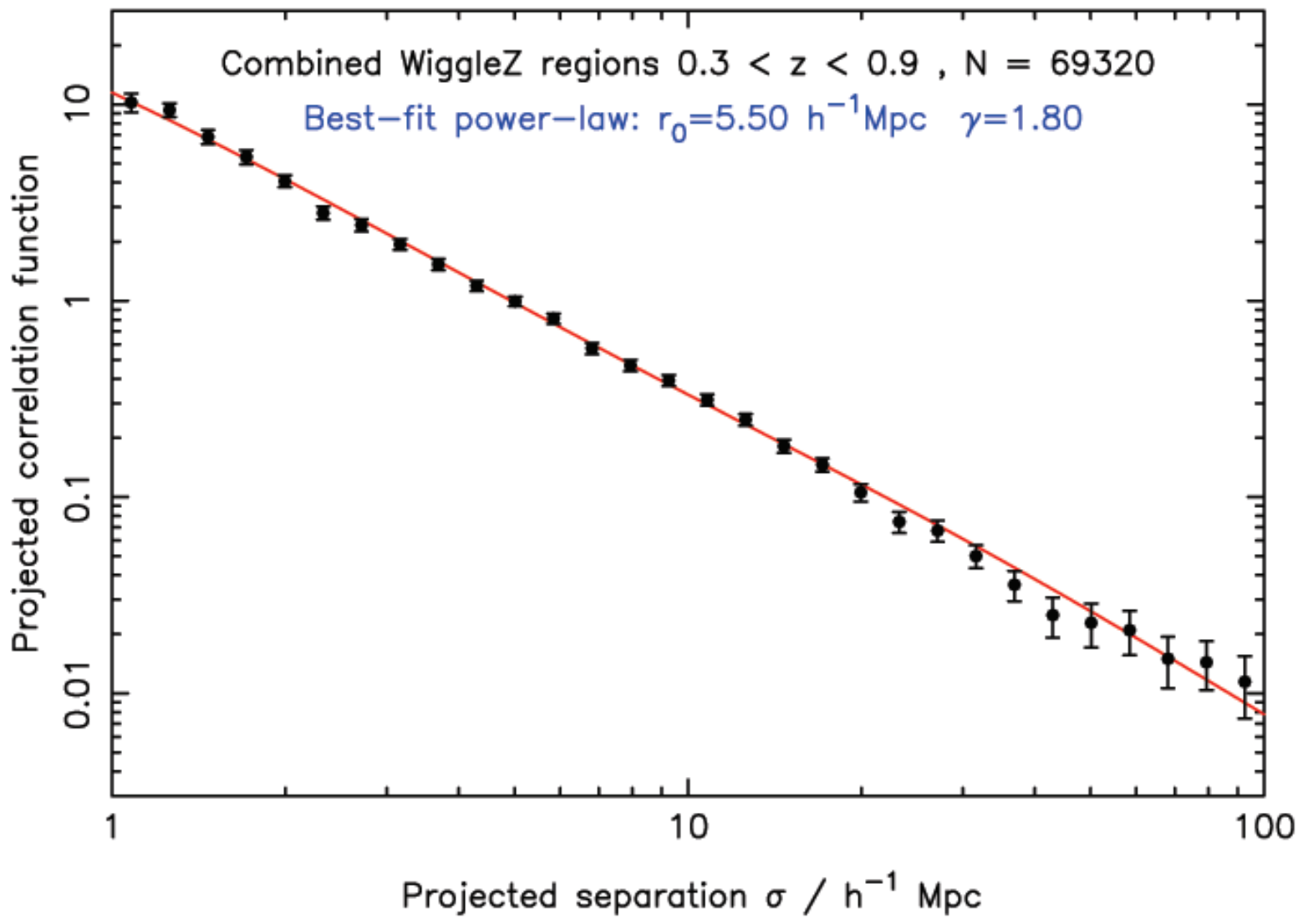

Figure 3: The projected correlation function of WiggleZ galaxies as a function of separation σ measured perpendicular to the line-of-sight.

10 $M_{\odot}$ $yr^{-1}$ at the average redshift, and exceeding 100 $M_{\odot}$ $yr^{-1}$ for the most distant objects.

Figure 2 displays a gallery of postage stamp images for some of our targets which have been serendipitously observed by the Hubble Space Telescope. These images typically reveal distorted morphologies or close companions indicative of galaxy interactions and mergers, processes which are known to trigger *starbursts* in galaxies.

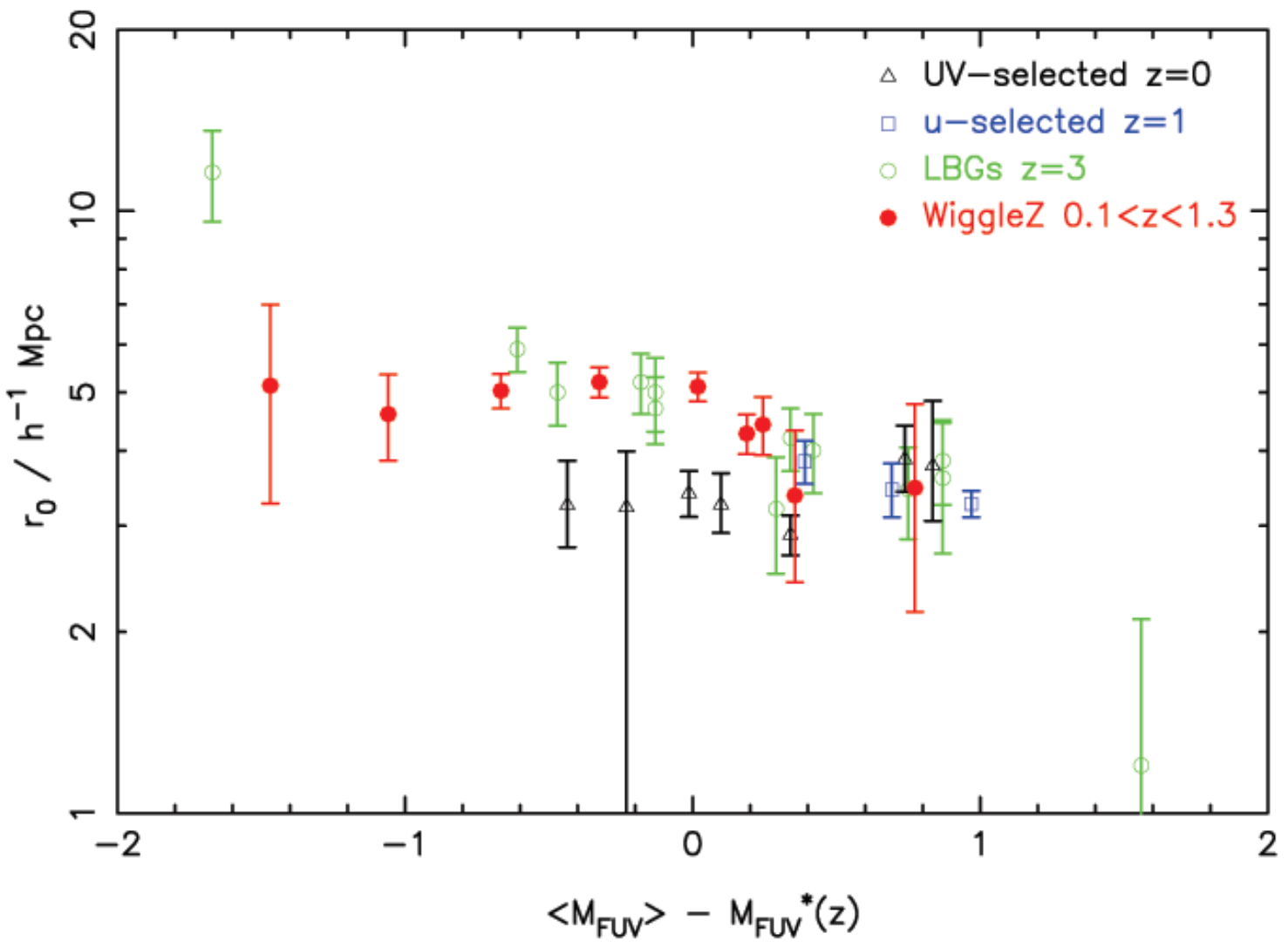

Figure 4: The dependence of the clustering length on UV luminosity for WiggleZ galaxies (solid red circles) compared to local UV-selected star-forming galaxies (black triangles) and high-redshift Lyman Break Galaxies (open green circles). The *x*-axis is the galaxy luminosity relative to the characteristic absolute magnitude *M**; brighter luminosities fall to the left of the diagram.

**The environments of the WiggleZ galaxies**

What is the local environment of WiggleZ galaxies: do they inhabit clusters, voids, or intermediate group environments? This question can be addressed by measuring the “clumpiness” of the WiggleZ galaxy distribution using a statistic known as the *projected correlation function,* which counts the number of pairs of close neighbours within the sample as a function of separation. This measurement is displayed in Figure 3; the data delineate a clean power law over a wide range of scales from 1 to 100 $h^{-1}$ Mpc (where $h = H_0/(100$ km $s^{-1}$ $Mpc^{-1})$ is the Hubble parameter). If we describe this power law using the formula $(r_0/r)^{\gamma}$, then the amplitude $r_0$ is known as the *clustering length* and describes the density of the local environment of the galaxies. We find values for these parameters of $r_0 \sim 5h^{-1}$ Mpc and $\gamma \sim 1.8$.

It is interesting to compare these results with those obtained for other classes of galaxy. Studies of local star-forming galaxies, such as performed by the 2dFGRS, find a significantly lower clustering length of $r_0 \sim 3h^{-1}$ Mpc. This suggests that the WiggleZ galaxies inhabit somewhat denser environments than local star-forming galaxies. This result is consistent with the suggestion that WiggleZ galaxies are “active

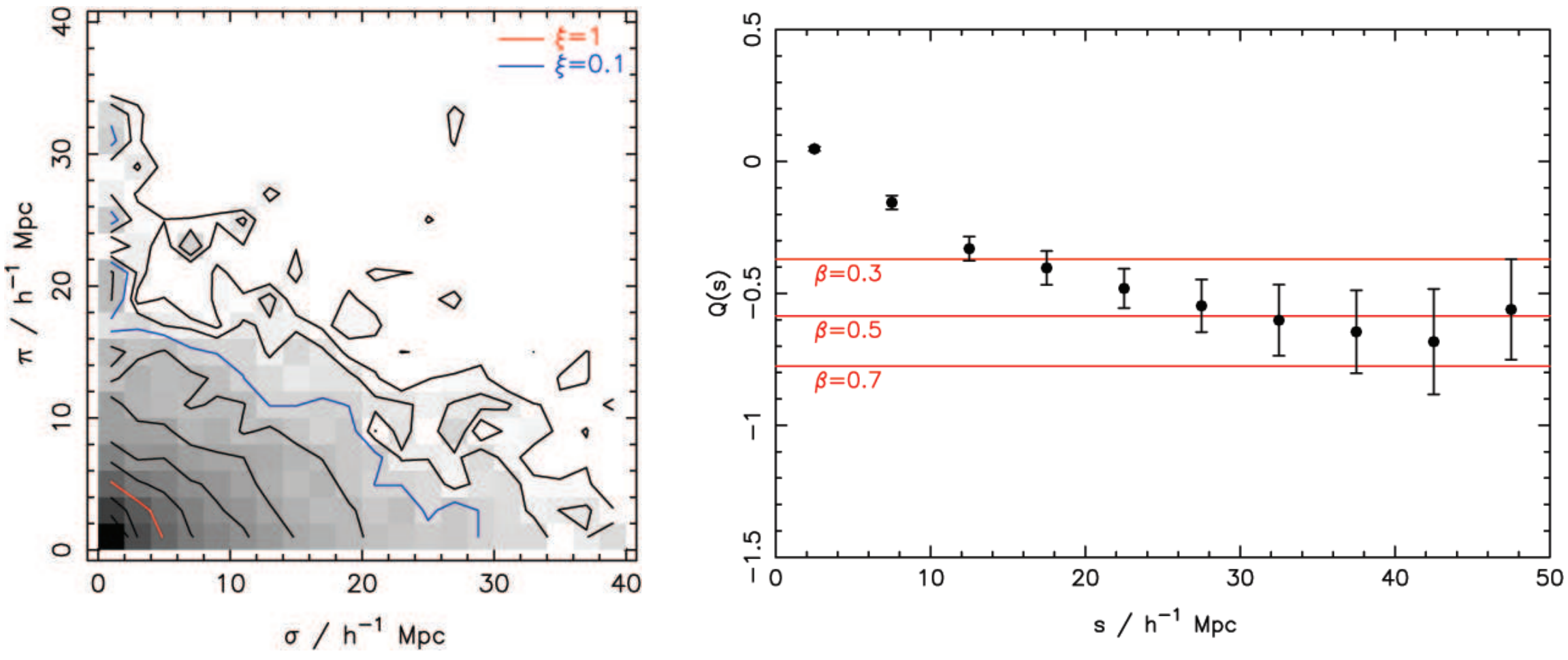

Left – Figure 5: The correlation function split by separations (σ,π) defined perpendicular and parallel to the line-of-sight (represented by both grey scale and contours). The deviations from circularity encode the imprint of the coherent infall of galaxies into clusters and superclusters, which traces the rate of growth of structure. The red (blue) contours correspond to correlation function values ξ=1 (0.1).
Right – Figure 6: The quadrupole moment of the WiggleZ correlation function, which quantifies the deviation of the contours of Figure 5 from circularity. On large scales > $20h^{-1}$ Mpc, this quantity converges to a constant which depends on the value of the redshift-space distortion parameter β (predictions for different values of β are shown on the plot). The value of β is in turn linked to the growth rate of cosmic structure.

starbursts" produced by mergers or interactions in denser group-scale environments at high redshifts. However, the clustering length for WiggleZ galaxies is lower than that found for Luminous Red Galaxies ($r_0 \sim 8h^{-1}$ Mpc) or for galaxy clusters ($r_0 \sim 15h^{-1}$ Mpc). This is unsurprising because star formation activity is known to be suppressed in the densest environments by mechanisms which remove the gas supply from galaxies.

Figure 4 analyses the dependence of the WiggleZ clustering length on the galaxy luminosity measured in the ultra-violet. Results for various other UV-selected galaxy samples are shown. At faint luminosities (on the right of the diagram) the WiggleZ galaxies have a similar clustering strength to local star-forming samples. However, at brighter luminosities (moving left in the diagram) the clustering length increases with luminosity in a manner similar to that observed for Lyman break galaxies (LBGs) at higher redshifts. This suggests that the nature of the WiggleZ sample, selected in part by requiring a Lyman break in the UV, is similar to higher-redshift LBGs. This is again consistent with a hypothesis that WiggleZ galaxies are produced by an active mode of star formation driven by galaxy mergers and interactions.

**Redshift-space distortions**

Galaxies possess "peculiar velocities" through space in addition to their overall motion in the Hubble expansion. Outside the cores of clusters, these peculiar velocities are produced by the coherent flow of galaxies into clusters through the action of gravity. They produce small changes in the measured redshifts of the galaxies which amount to a "flattening" of the large-scale appearance of clusters. This flattening arises because, seen from our vantage point, galaxies lying in front of clusters are falling away from us into the cluster increasing their redshifts, whereas galaxies lying behind clusters are falling towards us into the cluster decreasing their redshifts. This distinctive signature can be measured in our data set by plotting the galaxy correlations as a 2D function of redshift-space separation parallel and perpendicular to the line-of-sight, as shown in Figure 5.

In Figure 6 we extract the *quadrupole moment* of this 2D distribution as a function of scale and compare the results to models corresponding to different values of the redshift-space distortion parameter β. This is a work in progress, the eventual goal being to map out the growth rate of cosmic structure across our sample from $z$=1 to $z$=0, which is a powerful test of models of dark energy.

**The galaxy power spectrum**

A powerful method of describing the large-scale pattern of galaxy clustering is to break up the distribution into its Fourier modes. The relative Fourier amplitudes of large-wavelength fluctuations, from 10 to 1000 $h^{-1}$ Mpc in scale, can be used to perform a clean measurement of the cosmological parameters. This is because these large-scale fluctuations are generated by interactions between atoms (baryons), dark matter and radiation in the very early Universe, prior to the generation of the Cosmic Microwave Background (CMB) radiation at redshift $z$~1000, processes which can be modelled extremely accurately. On large scales exceeding 10 $h^{-1}$ Mpc, the resulting spectrum of density fluctuations grows in a simple, linear fashion between $z$=1000 and $z$=0. Therefore, mapping the galaxy distribution today allows us to infer the relative fractions of atoms, dark matter and dark energy in the Universe.

This is achieved in practice by measuring the *galaxy power spectrum,* which describes the strength of the Fourier amplitudes as a function of Fourier scale $k$ (low values of $k$ indicating large scales). The power spectrum measured for WiggleZ galaxies is plotted in Figure 7, together with a model prediction for a Universe which today contains fractions (4%, 21%, 75%) of its present energy density in the form of baryons, dark matter and dark energy, respectively. (These fractions were chosen because they also produce a good description of fluctuations in the CMB). The model produces a good fit to our data, constituting a valuable consistency check of cosmological data which demonstrates that galaxies grew from small density seeds that were present in the CMB.

We can also fit cosmological models to the WiggleZ power spectrum in order to measure the dark matter and baryonic fractions. The result of this measurement is shown in Figure 8 and is fully consistent with the consensus that firstly, the great majority of matter exists as non-baryonic, cold dark matter and secondly, that all forms of matter taken together comprise only ~30% of the energy of the Universe.

## Summary

The WiggleZ project is scheduled to complete observations in Semester 2010A, at which point we will be able to measure the BAO scale, allowing us to make an accurate and independent test of the constant dark energy model [11]. In the meantime, we have already collected the largest sample of emission-line galaxies at these redshifts ever obtained. We have summarised our initial scientific results from the sample in this article. Later this year we will make a public release of the spectroscopic data obtained so far for other groups to use.

## References

[1] Brumfiel G., 2003, Nature, 422, 108
[2] Efstathiou G., Sutherland W.J., Maddox S.J., 1990, Nature, 348, 705
[3] Riess A.G., et al., 1998, AJ, 116, 1009
[4] Perlmutter S., et al., 1999, ApJ, 517, 565
[5] Deffayet C., et al., 2002, PRD, 65, 044023
[6] Bennett C., et al., 2003, ApJS, 148, 1
[7] Cole S., et al., 2005, MNRAS, 362, 505
[8] Eisenstein D.J., et al., 2005, ApJ, 633, 560
[9] Blake C., Glazebrook K., 2003, ApJ, 594, 665
[10] Martin D.C., et al., 2005 ApJ, 619, L1
[11] Blake, C., et al., 2009, MNRAS, in press

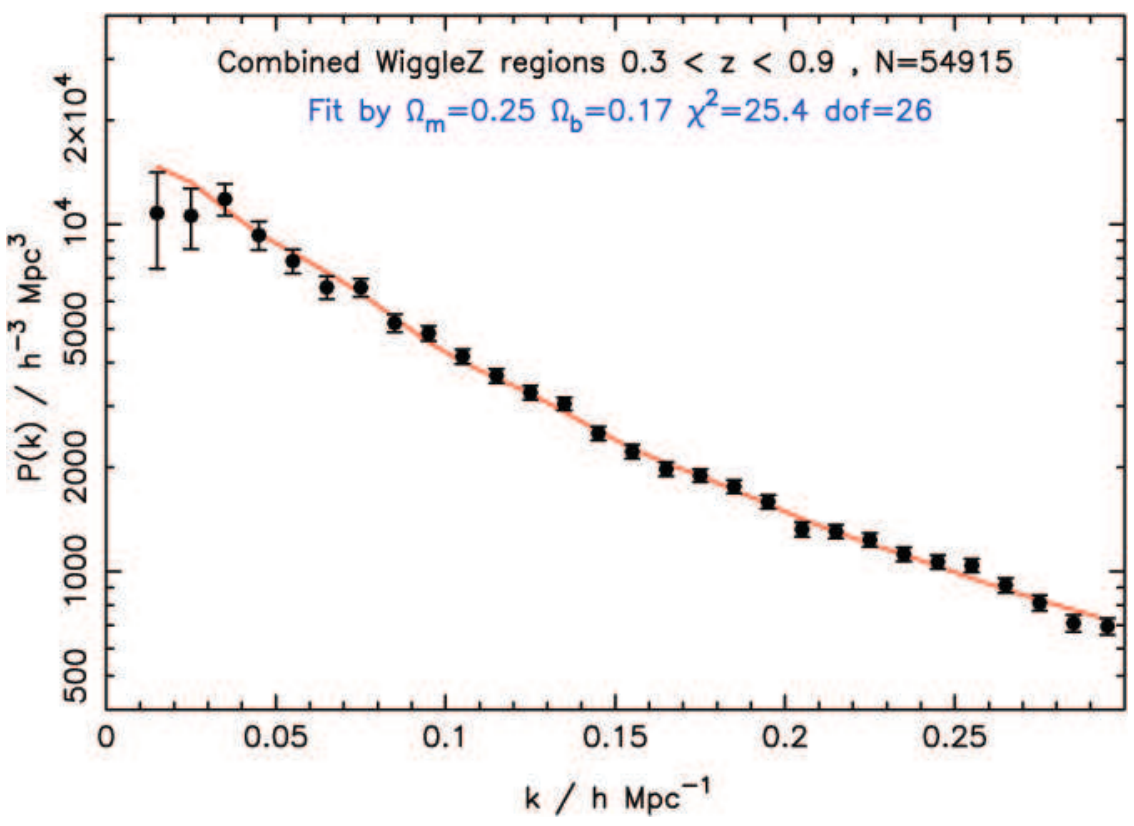

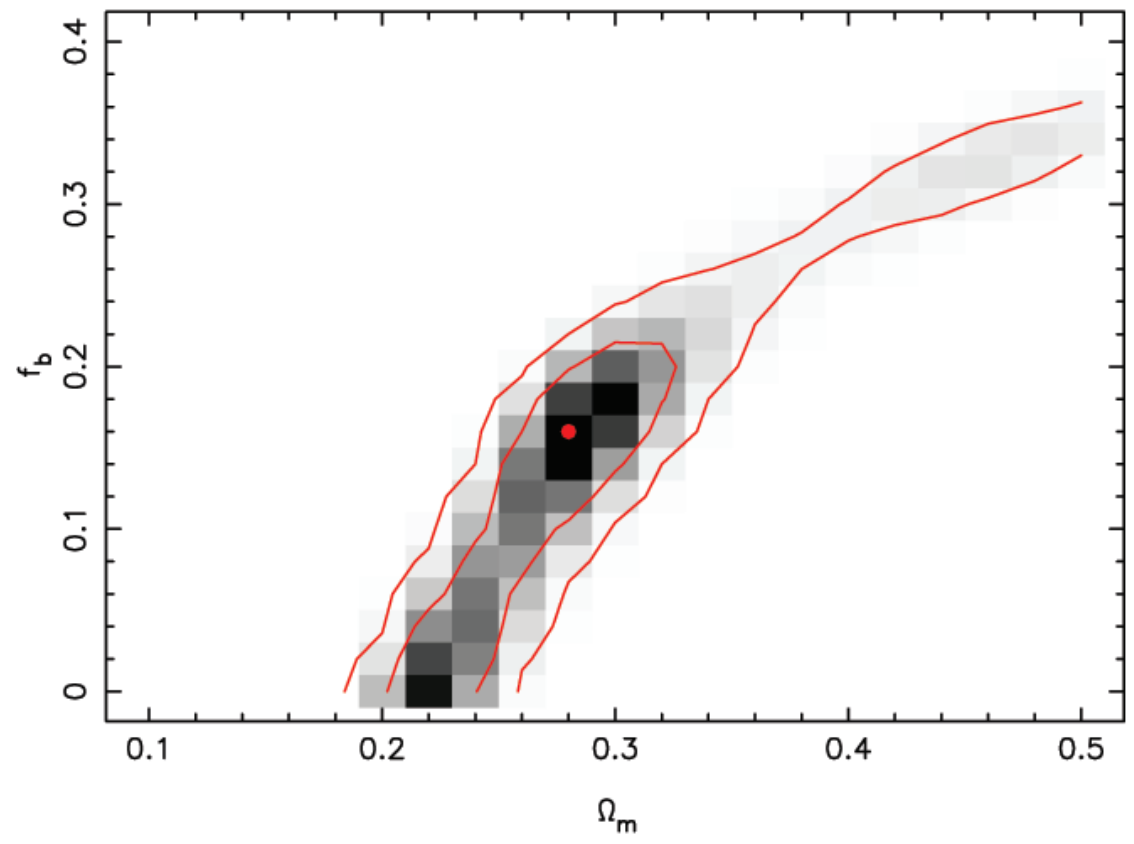

Left – Figure 7: The data points indicate the WiggleZ galaxy power spectrum obtained by combining measurements in the 5 most advanced survey regions, compared to a linear model power spectrum with standard cosmological parameters.
Right – Figure 8: Probability contours obtained by fitting values of the total matter density $\Omega_m$ (on the $x$-axis) and baryonic fraction $f_b$ (on the $y$-axis) to the WiggleZ galaxy power spectrum. 1-$\sigma$ and 2-$\sigma$ contours are shown, enclosing 68% and 95% of the total probability, respectively.

# IRIS2 UNWINDS A SPIRAL GALAXY?

Stuart Ryder (AAO), Petri Väisänen (South African Astronomical Observatory), Seppo Mattila and Jari Kotilainen (Univ. of Turku)

## Introduction

As recently as half a century ago, astronomers were still vigorously debating whether all spiral galaxies had trailing arms consistent with differential rotation in their inner parts, or if in fact some galaxies had “leading arms” which curve in the opposite sense to the disk rotation. While de Vaucouleurs (1959) firmly believed in trailing arms, Lindblad (1951) argued that he could see some cases of leading arm spiral galaxies.

In theory at least, determining whether spiral arms are leading or trailing ought to be simple enough. A longslit spectrum across the major axis of the inclined galactic disk will give an unambiguous indication of which side of the galaxy is approaching, and which side is receding from, us. Knowing which side of the disk is the near side to us and which side is farther away will then tell us immediately whether the arms are trailing or leading the rotation. Unfortunately there are few unequivocal methods for working out the viewing geometry of a galactic disk. Perhaps the most intuitive is to look for the obscuring effect of dust lanes in highly-inclined systems. Close examination of high-quality images of galaxies such as the famous “Sombrero” galaxy M104 (http://www.aao.gov.au/images/captions/aat100.html) or NGC 7331 (http://antwrp.gsfc.nasa.gov/apod/ap081022.html) leaves one with the very clear impression that the side of the disk most silhouetted by dust lanes must be the one nearest to us. If this is generally the case, then IRIS2 may have helped reveal the best case yet for a leading arm spiral galaxy, albeit one undergoing an interaction with a smaller companion.

### The Curious Case of IRAS 18293-3413

Since 2004, our collaboration has been using adaptive optics systems on the ESO VLT and Gemini North 8 metre telescopes in an effort to find previously hidden, highly-extinguished core-collapse supernovae in Luminous Infrared Galaxies (LIRGs, for which $L_{IR} > 10^{11}\ L_{\odot}$), whose prodigious star formation rates ought to make them ideal hunting grounds for supernovae. Our first success came with the discovery of SN 2004ip in the LIRG IRAS 18293-3413 (Mattila et al. 2007), followed by SN 2008cs in IRAS 17138-1017 (Kankare et al. 2008). An added legacy of this supernova search campaign is the set of high resolution images accumulated over multiple epochs of the LIRGs themselves, which can show us how and where star formation is being triggered (e.g., Väisänen et al. 2008a). The images of IRAS 18293-3413 obtained with NAOS-CONICA on the VLT and shown in Figure 1 are a case in point. What appears to be a dusty mess with flocculent spiral structure in the optical with the HST reveals itself to have “grand design” spiral arms in the near-infrared, as well as a possible companion galaxy 15" to the northwest.

In order to ascertain whether the putative companion has a similar redshift to IRAS 18293-3413, we obtained longslit spectra covering the *J*, *H*, and *K*-bands of both objects simultaneously as marked in Figure 1 with IRIS2 on 2007 Sep 27 and 28. The rotation curves derived for various emission and absorption features are shown in Figure 2. We derive a redshift for the adjacent galaxy of 5960 ± 80 km s$^{-1}$, some 500 km s$^{-1}$ greater than IRAS 18293-3413, which means that if it is a companion

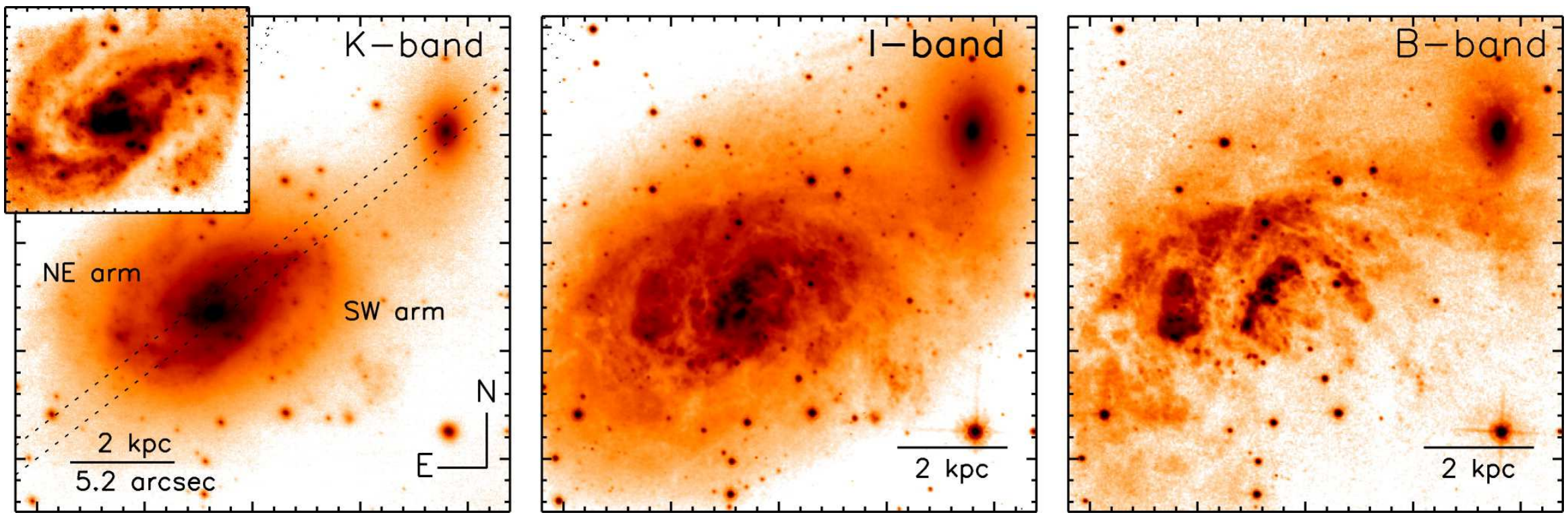

Figure 1: The LIRG IRAS 18293-3413 imaged in *K* with NAOS-CONICA on the VLT (left), and in *I* (middle) and *B* (right) with ACS on the HST. The inset at left is an unsharp-masked version of the *K* image which highlights the spiral structure, while the dashed lines in the left panel indicate the IRIS2 slit position which crosses the major axis of IRAS 18293-3413 as well as its companion.

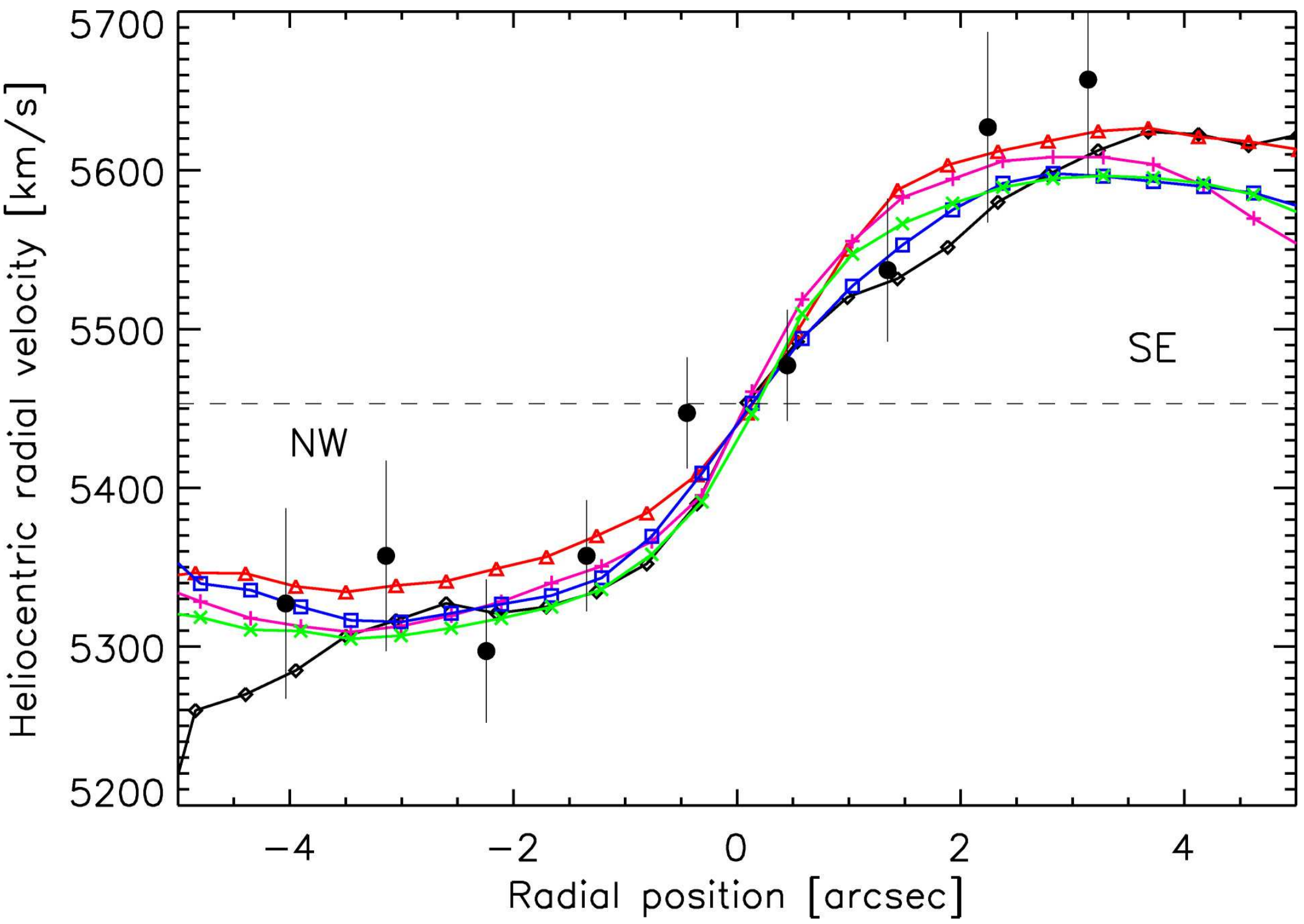

Figure 2: Line of sight velocity curves derived from IRIS2 spectra of the CO 2.3 μm bands (**black** filled circles); [Fe II] 1.257 μm line (black diamonds); Pa α 1.876 μm (**red** triangles); He I 2.059 μm (**pink** plus signs); $H_2$ 1–0 S(1) 2.122 μm (**blue** squares); and Br γ 2.166 μm (**green** crosses).

galaxy to the LIRG then it must be undergoing a high-speed, retrograde encounter.

Following the arguments outlined in the preceding section about the most heavily dust-obscured side of the galaxy being closest to us, one must conclude that the southwest side of the disk of IRAS 18293-3413 is in the foreground. Meanwhile Figure 2 shows that the northwest side of the disk is approaching us, so the galaxy is rotating clockwise even though the arms open out in the same direction. The inescapable conclusion is that *the spiral arms revealed in our near-IR adaptive optics images are leading spiral arms*, something which only became apparent from the IRIS2 spectra. While it is conceivable that the dust structure does not lie in the plane of the disk as assumed, and could perhaps be entrained in an AGN-driven wind out of the disk plane, we see no evidence for an AGN in the near-IR spectra.

**Why so rare?**

While most LIRGs appear to be the result of an interaction/merger between 2 or more gas-rich spirals (Väisänen et al. 2008a), IRAS 18293-3413 seems instead to consist of a small elliptical galaxy undergoing a first-time, high speed, retrograde encounter with a spiral galaxy. Numerical modelling by Thomasson et al. (1989) and by Byrd et al. (1993) does indeed predict that retrograde encounters with small companions can result in long-lived leading arm features, albeit only a single arm, and only when the dark matter halo outweighs the spiral itself. Since retrograde encounters ought to be just as frequent as prograde encounters, perhaps the extreme rarity of a double leading arm system like IRAS 18293-3413 indicates that such massive halos are also uncommon?

As adaptive optics imaging of LIRG systems accompanied by kinematic studies like this one are still few and far between, it may be that further examples of leading arm systems are out there waiting to be discovered. Although the discovery of supernovae is what drove our study originally, the serendipitous discovery of the best candidate yet for a genuine leading arm spiral is an unexpected bonus. For a full discussion of this work, see Väisänen et al. 2008b.

**References**

Byrd, G., Freeman, T., & Howard, S. 1993, AJ, 105, 477
de Vaucouleurs, G. 1958, ApJ, 127, 487
Kankare, E. et al. 2008, ApJL, 689, L97
Lindblad, B. 1951, PASP, 63, 133
Mattila, S., et al. 2007, ApJL, 659, L9
Thomasson, M., et al. 1989, A&A, 211, 25
Väisänen, P., et al. 2008a, MNRAS, 348, 886
Väisänen, P., et al. 2008b, ApJL, 689, L37

## UNVEILING THE DYNAMICS OF GALAXIES WITH SPIRAL

Andy Green, Karl Glazebrook, Greg Poole (Swinburne), Peter McGregor (ANU), Bob Abraham, Ivana Damjanov (Toronto), Pat McCarthy (OCIW), Rob Sharp (AAO)

The SPIRAL integral field spectrograph is proving to be an excellent instrument for observing local emission line galaxies for resolved 2D kinematic and chemical analysis. The dynamical evolution of galaxies, particularly spiral galaxies like the Milky Way, still presents challenges to $\Lambda$CDM cosmology. It is unclear whether the growth of large galaxies at high-redshift is mostly driven by major mergers or by in-situ rapid star-formation in early disks. In the local Universe there are still theoretical problems in explaining the dynamical scaling relations of disk galaxies.

Recently, IFU instruments on 8 metre class telescopes around the world have measured the 2D velocity fields of star-burst galaxies at $z$~2–3, painting a picture of complex activity. Because of their small apparent size, an IFU and adaptive optics are critical to understanding the dynamical state of these distant galaxies (Flores et al. 2006). Large galactic disks have been seen as early as just 3 billion years after the big bang, but these disks are often very different from those observed in the present universe (Genzel et al. 2006). Moreover, more than half of all observed star forming galaxies at $z > 1$ do not appear to be disks, while star formation commonly indicates the presence of a disk in the local universe (Law et al. 2007). Our group is pursuing dynamics of mass selected samples at $1<z<2$ using Gemini and Keck laser AO and it has come to our attention that the local datasets are insufficient for interpreting the high-redshift Universe.

Traditionally, our understanding of galaxy dynamics in the local universe has been from unresolved spectroscopy (e.g. Tully and Fisher 1977) or long-slit spectroscopy, providing only one-dimensional spatially resolved information across the galaxy (e.g. Courteau 1997). Unfortunately little work has been done with spatially resolved dynamical information of local galaxies. Notable exceptions are the GHASP survey (e.g. Spano et al. 2008), which uses a scanning Fabry-Perot and photon counting techniques, and the SAURON project (Bacon 2001), which uses a low dispersion IFU. However, there is no comparable dataset, in terms of resolution and selection, in the local Universe that can guide us in interpreting the high redshift Universe.

**Enter SPIRAL and the DYNAMO Project**

To address this problem, the DYNAMO project is now using SPIRAL to survey a representative sample of star forming galaxies in the local universe selected from the Sloan Digital Sky Survey in order to bolster our understanding of local galaxy dynamics. The project aims specifically at developing a better understanding of local galaxies such that meaningful comparisons between present and past star forming galaxy dynamics can be made. Galaxies at high redshift tend to be either "kinematically disturbed" or "hot disks" (Flores 2006, Law 2007). However, local galaxies have not been classified by their kinematical state in the same way. DYNAMO will characterise evolution in the fraction of galaxies in various dynamical classes, which can be compared with simulations. By artificially redshifting data from local galaxies to simulate high-redshift observations, we can check that those observations are robust.

Furthermore, having the full dynamical information about a galaxy enables rotation velocity measurements of disks that are robust against small disturbances. This substructure may be a major source of scatter in the Tully-Fisher relation, and can be quantified and removed from IFU data. Also, the number density of galaxies vs the disk circular velocity (a form of the galaxy mass function) allows us to test predictions of cold dark matter cosmology. A particular issue is the ratio of disk rotation velocity to the halo rotation velocity, which is difficult to account for in a $\Lambda$CDM cosmology (the 'angular momentum problem', Sommer-Larsen & Dolgov 2001).

Finally, because SPIRAL's field of view is larger than most of the galaxies we observe, and because of the tuneable spectral coverage and resolution provided by AAOmega, we will be able to measure galaxy properties unbiased by the aperture correction present in Sloan Digital Sky Survey spectra, and other fibre fed spectral surveys. Aperture corrections are often large, and could easily dominate the uncertainty in the measurement of many standard galaxy properties (Brinchmann et al. 2004).

**Preliminary Results**

Our experimental design focuses on the H$\alpha$ emission line to recover kinematics, the same tool being used to study $z$~2 galaxies. H$\alpha$ traces HII regions in galaxies which are the sites of current star formation. Several velocity maps, which measure the line of sight velocity of these regions in the galaxy relative to the system's total velocity, are shown in Figure 1. Our first observing run in July 2008 sampled some of the most luminous star forming galaxies in the Sloan Digital Sky Survey. Further observations will move down the Hα luminosity function to survey a more complete sample of these local galaxies.

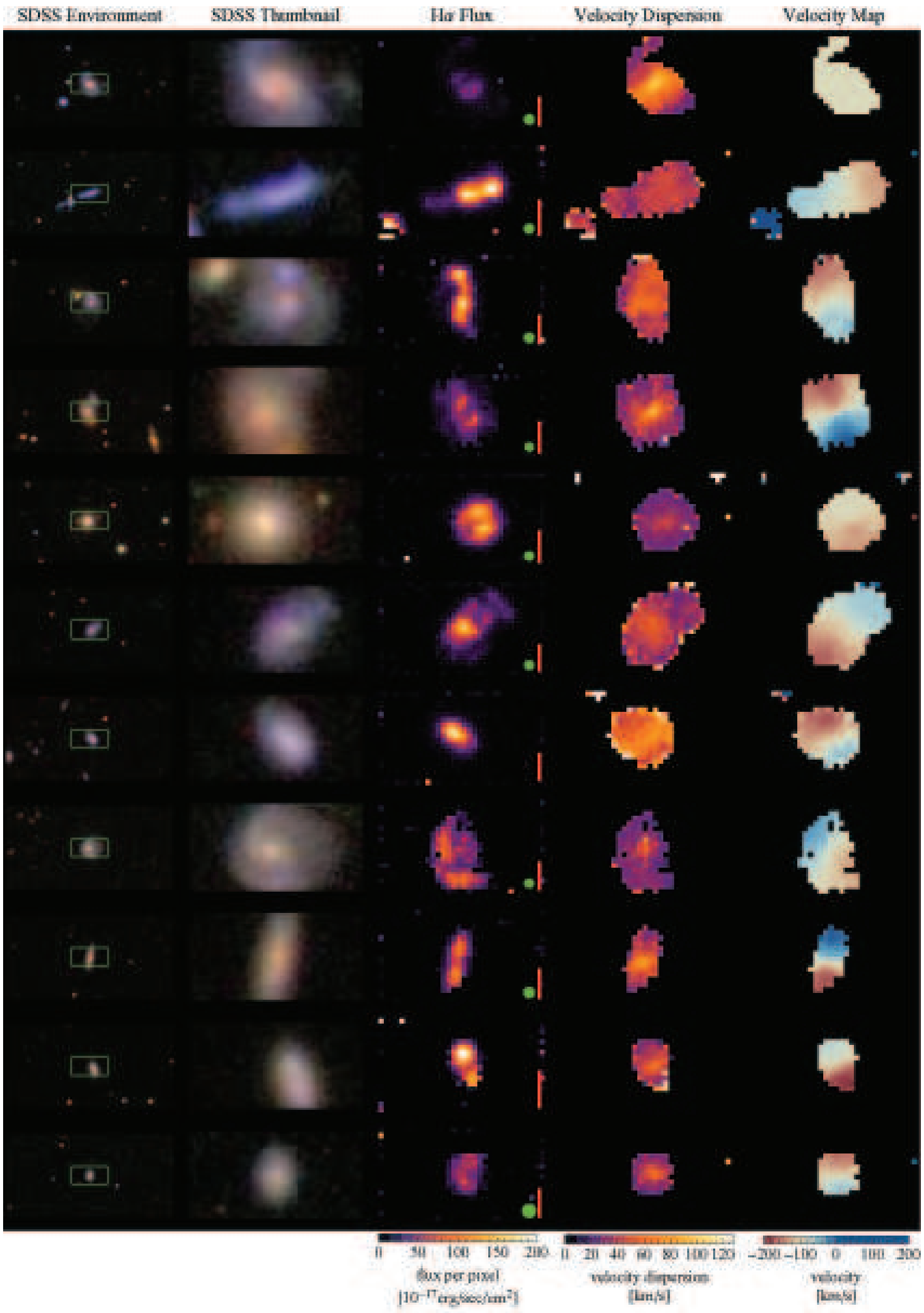

Figure 1: Sample of results from our first observing run, July 2008. All images present the same field of view. The green disk shows the FWHM of the seeing for our observations, while the red bar shows the extent of 5 kiloparsecs at the galaxy's redshift. Several of these galaxies have important kinematic substructure which long slit observations would not have characterised. We are able to recover velocity information over much of the galaxy's visible light, eliminating aperture effects.

Although we are still in the early stages of our analysis of these galaxies, the wealth of information is apparent in Figure 1 and in the centrefold. Comparing the broadband light to the flux of Hα shows where star formation is occurring relative to the rest of the galaxy's stars, and might suggest external influences. Through various techniques, including Kinemetry (Krajnovic et al. 2006), the velocity fields of mergers can be readily separated from quiescent disks. Comparing optical morphologies to kinematic morphologies will advance our understanding of other galaxies. And finally, careful comparison of dynamical mass to stellar mass may answer long-standing questions about how dark matter is distributed within galaxy halos.

# The Dynamics of Newly Assembled Mass

Andy Green, Karl Glazebrook, Greg Poole (Swinburne), Peter McGregor (MSO), Roberto Abraham, Ivana Damjanov (Toronto), Pat McCarthy (OCIW), Rob Sharp (AAO)

DYNAMO will help unlock the secrets of galaxy formation by observing a variety of local and distant emission line galaxies with SPIRAL and other integral field spectrographs. The Sloan Digital Sky Survey's wealth of available data makes it the ideal source for our SPIRAL targets. Below are portraits randomly selected from our galaxies, with the SPIRAL F.O.V. overlaid. Targets for the project have been chosen based on Hα luminosity and other observational constraints such as sky line contamination, as well as background data available.

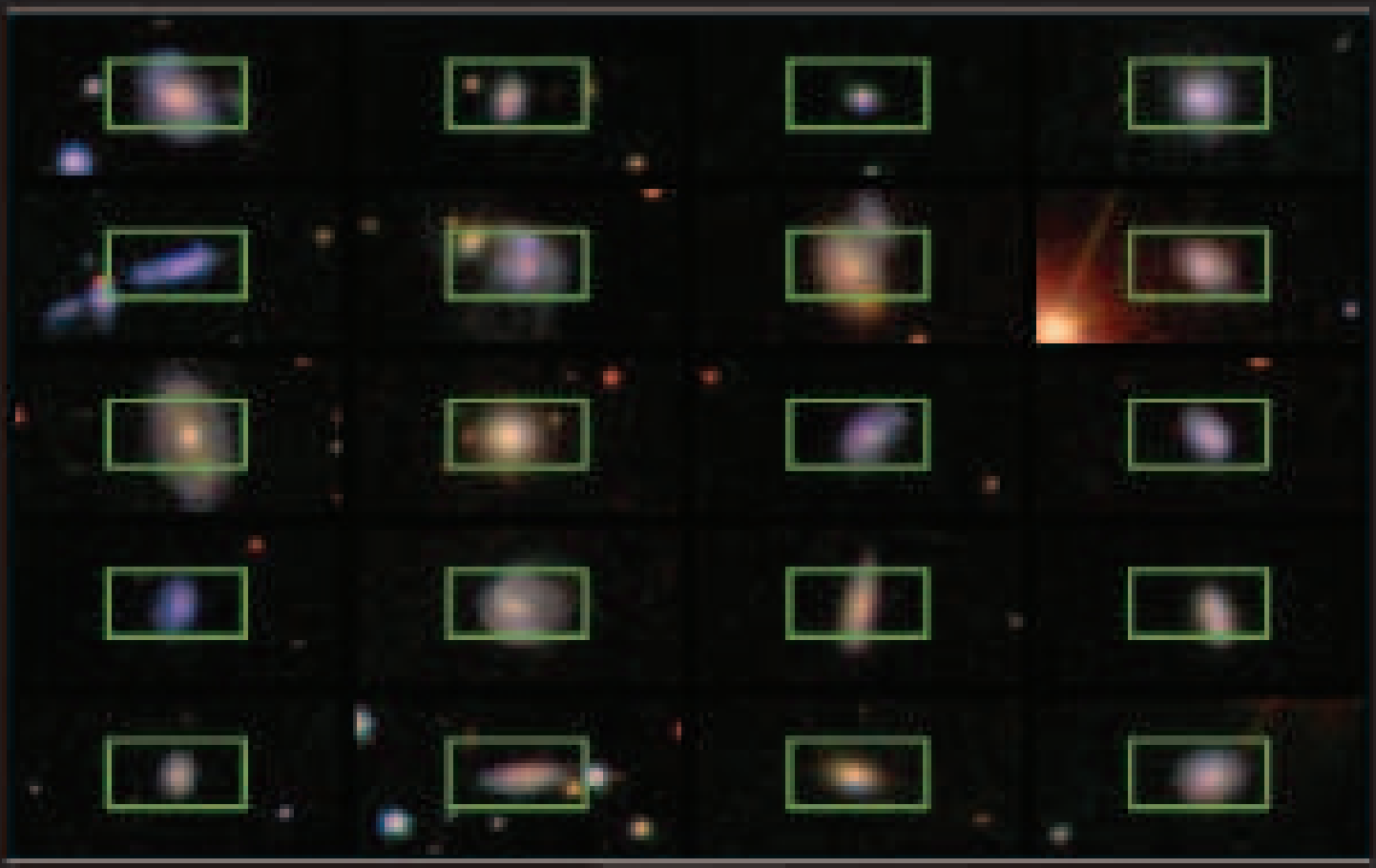

## Aperture E

SPIRALs confi
allows us to t
lines important
important galax
as metallicity a
The field of vi
much larger tha
(the red circle),
which often dom
unimportant in D

## IFU Data Cube

The key to our observations is the SPIRAL spectrograph, which creates a 3D data cube where two axes are spatial, and one is spectral. Combining the features of both a imager and a spectrograph into a single instrument, the integral field spectrograph provides the wealth of information shown here.

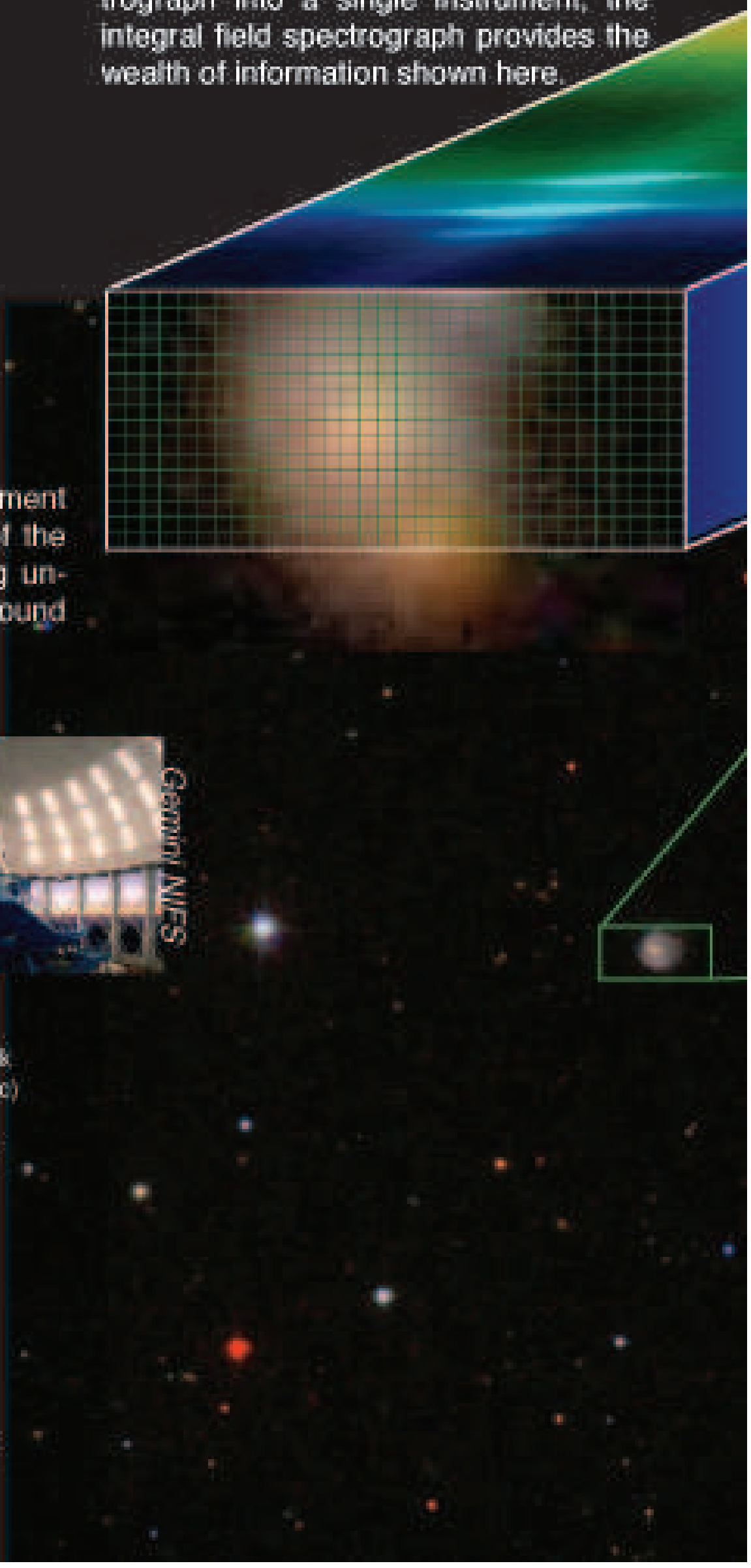

## Other IFUs

IFUs on Keck and Gemini coupled with Laser Adaptive Optics can observe galaxy dynamics during the peak of galaxy formation. When combined with SPIRAL data, these observations will help unravel the mystery of spiral galaxy formation. This data will compliment existing IFU observations of the high redshift universe being undertaken by astronomers around the world.

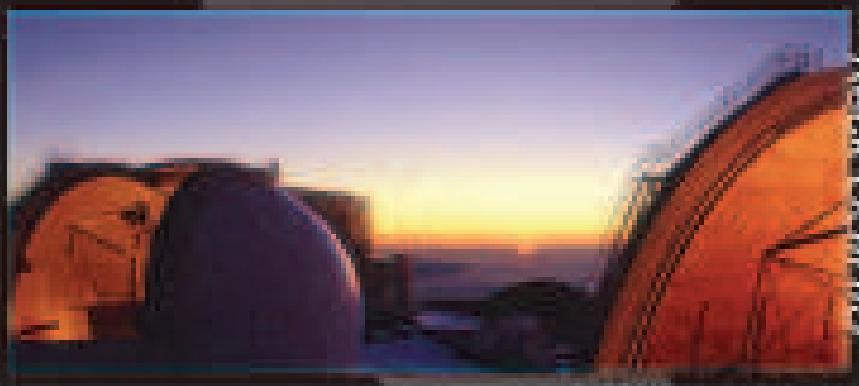

Keck OSIRIS

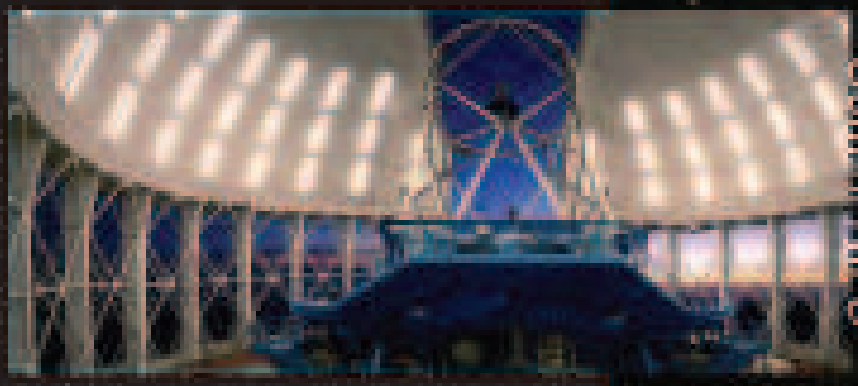

Gemini NIFS

## Artificial Redshifting

By simulating the effects of increasing distance (redshift) on our local targets, we can understand what these galaxies would look like at high redshift. These artificially redshifted targets can then be compared to real observations like this one from Gemini. Testing the methods used to interpret high redshift galaxies is a key goal of our project.

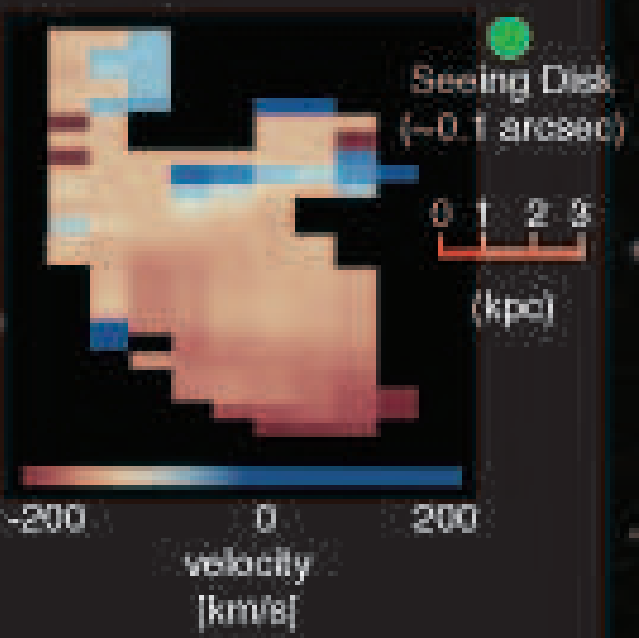

# ssive Objects

## e Effects, Chemistry

configurable spectral range to target specific emission tant to understanding many alaxy properties in 2D, such ity and star formation rate. of view (green squares) is r than that of an Sloan fiber cle), so aperture corrections dominate Sloan results, are t in DYNAMO.

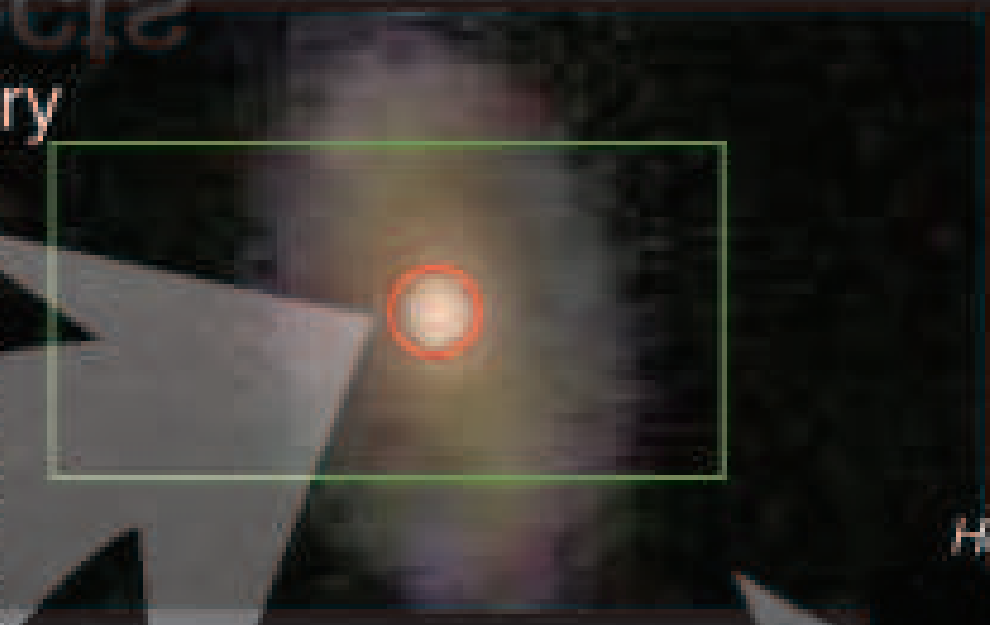

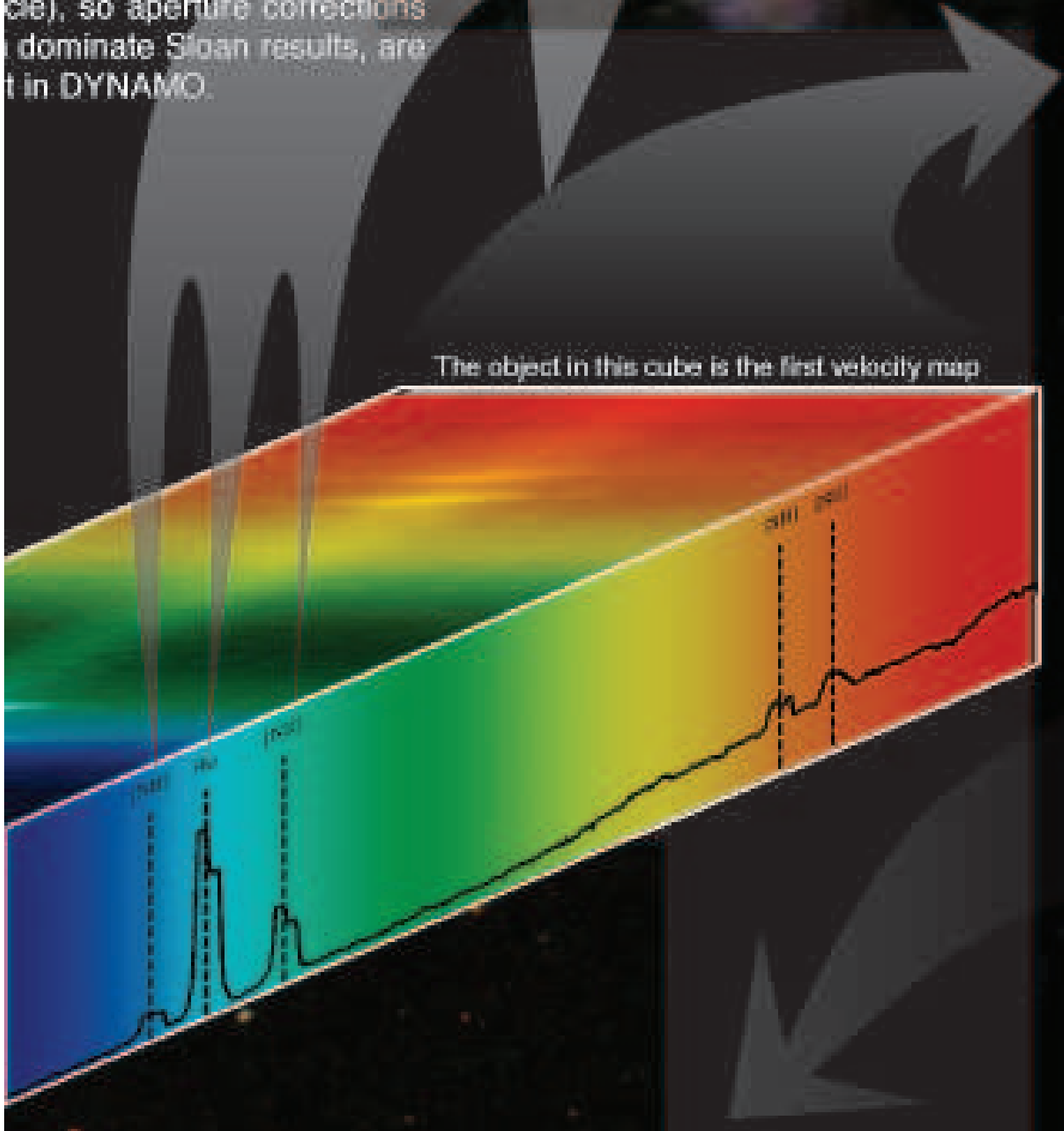

## Velocity Maps

By measuring the redshift of the Hα line in each spaxel, the relative velocity and velocity dispersion of HII clouds in the galaxy can be measured. For ordered disks, the galaxy circular velocity function can be measured, and substructure quantified. The velocity information also allows separation of ordered disks from mergers or other disordered systems.

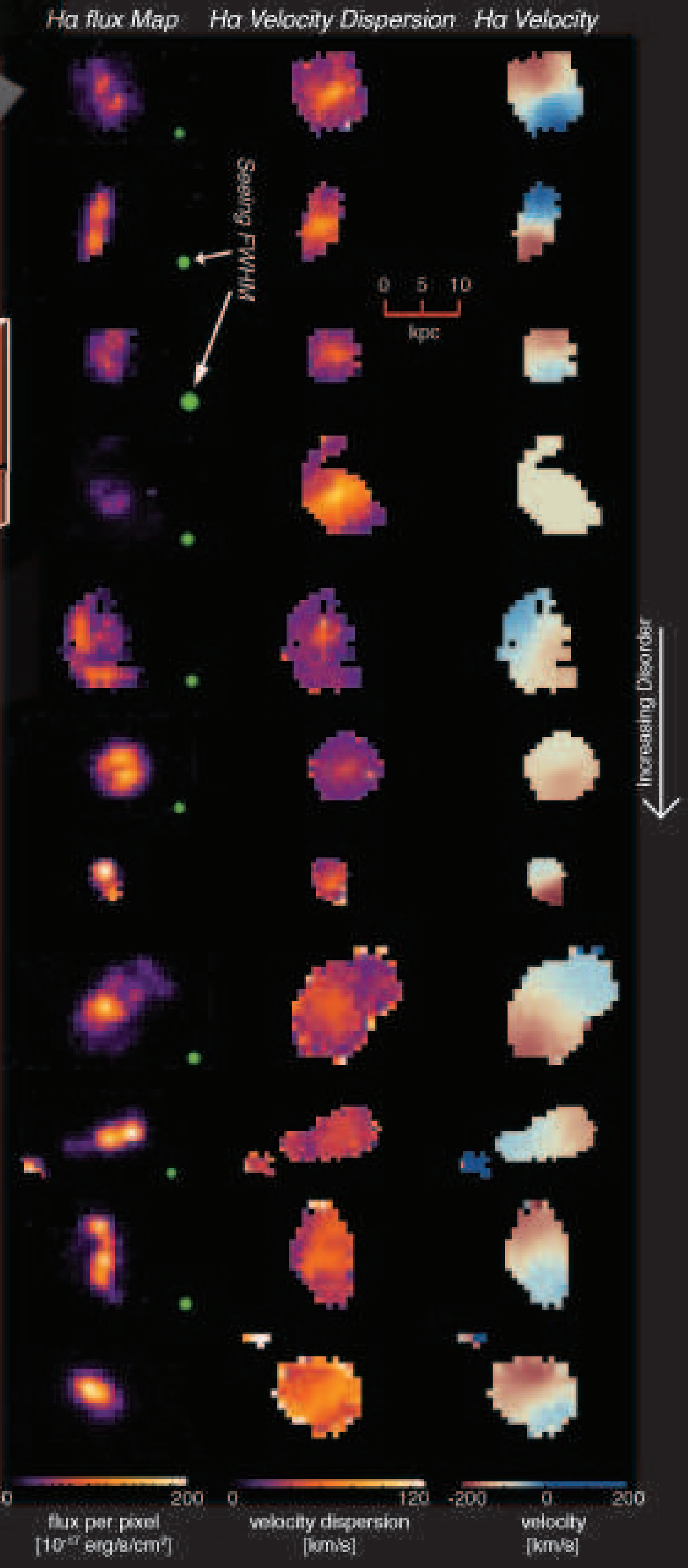

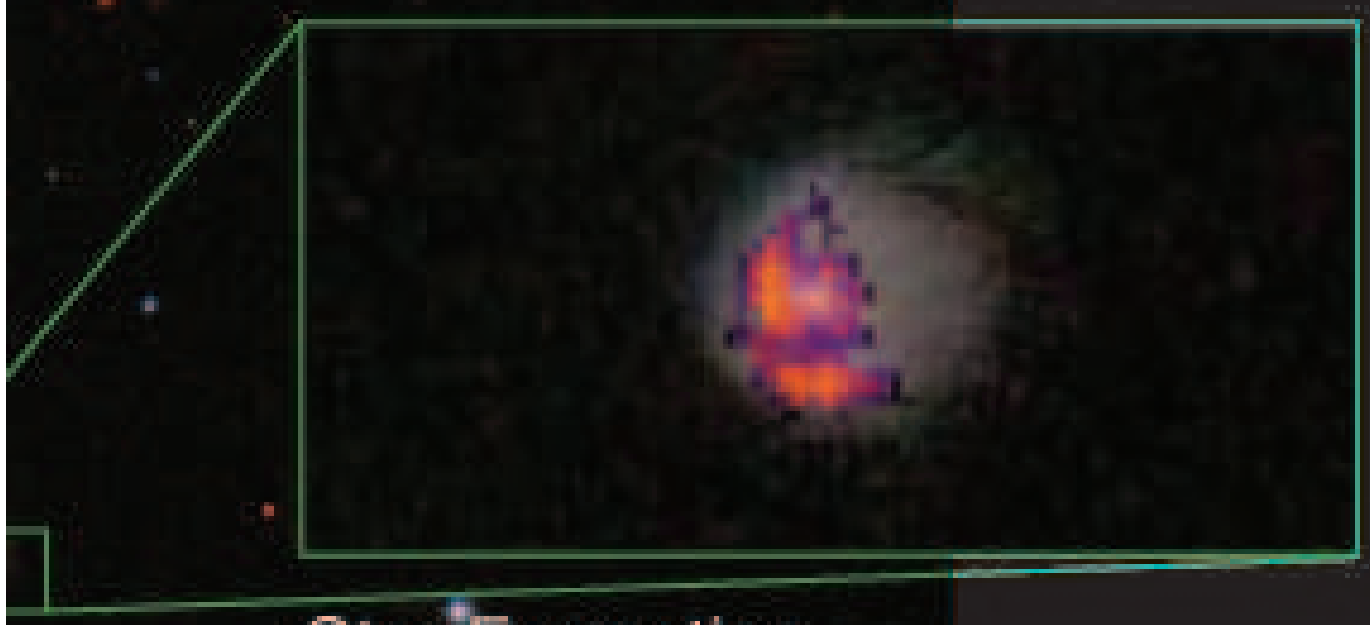

## Star Formation

The strength of the Hα line reflects the star formation rate in each part of the galaxy. Here, Hα has been superimposed on the Sloan thumbnail. This galaxy shows star formation on only one side, which might indicate some environmental effect such as ram pressure stripping. However, the Sloan image of the area reveals no obvious nearby clusters or other interesting environment.

## SPIRAL: A Triumph of AAOmega

SPIRAL is a front end which provides Integral Field capability to the AAOmega spectrograph. The 22.4 x 11.2 arcsecond field is sampled at the Cassegrain focus with 0.7" spatial pixels (spaxels) fed by fibre into AAOmega. This provides excellent field of view and resolution for seeing limited (generally 1.0"–1.5") observations of $z$~0.1 galaxies. This is the only IFU with the field of view and spectral resolution options matched to SDSS available to Australian scientists. By an interesting and useful coincidence the spatial resolution (in kpc) of galaxies at $z$~0.1 observed with SPIRAL are identical to those of typical laser AO IFU systems being used at $z$>1.

The SPIRAL system is a reincarnation of the earlier Integral Field Unit (Lee D., AAO Newsletter 93, p.8) coupled to the AAOmega spectrograph and taking advantage of commonalities with 2dF. This relatively low cost instrument is able to take advantage of the capabilities, infrastructure and data reduction suite (the output data format is identical) that has made the AAOmega spectrograph so popular among astronomers.

### Measurement of SPIRAL's throughput

Over the course of our five night run, we observed several photometric standards – enabling us to measure the instrument's absolute throughput in production using the 1700I VPH grating on the red arm of AAOmega. We present our results here as a service to the community as earlier measurements were made before the final honing of the instrument.

The raw frames were reduced using 2dfdr in the same manner as our science frames using Rob Sharp's Optimal Extraction routine for SPIRAL. We measured the throughput spectrum by summing the pixels spatially around the star in the IFU datacube, and then dividing this spectrum by that expected for the chosen standard star. The raw spectrum (in electrons detected/sec) is shown with the throughput spectrum in Figure 2.

The throughput, peaking at 12%, falls only slightly short of that predicted by the SPIRAL exposure time calculator, and is impressive considering the fibre-fed, Coudé room spectrograph and conservative budget of SPIRAL. The throughput and the relative simplicity of the SPIRAL/AAOmega design makes data collection and reduction surprisingly easy and objective. Many of the DYNAMO project's main goals would not be possible without this instrument.

### References

Bacon, R. et al. 2001, MNRAS, 326, 23
Brinchmann, J. et al. 2004, MNRAS, 351, 1151
Courteau, S. et al., 1997, AJ, 114, 2402
Flores, J.C. et al., 2006, A&A, 455, 107
Genzel, R. et al. 2006, Nature, 442, 786
Krajnovic, D. et al. 2006, MNRAS, 366, 787
Law, D.R. et al. 2007, ApJL, 656, 1
Sommer-Larsen, J. & Dolgov, A. 2001, ApJ, 551, 608
Spano, M. et al. 2008, MNRAS, 383, 297
Tully, R.B. & Fisher, J.R. 1977, A&A, 54, 661

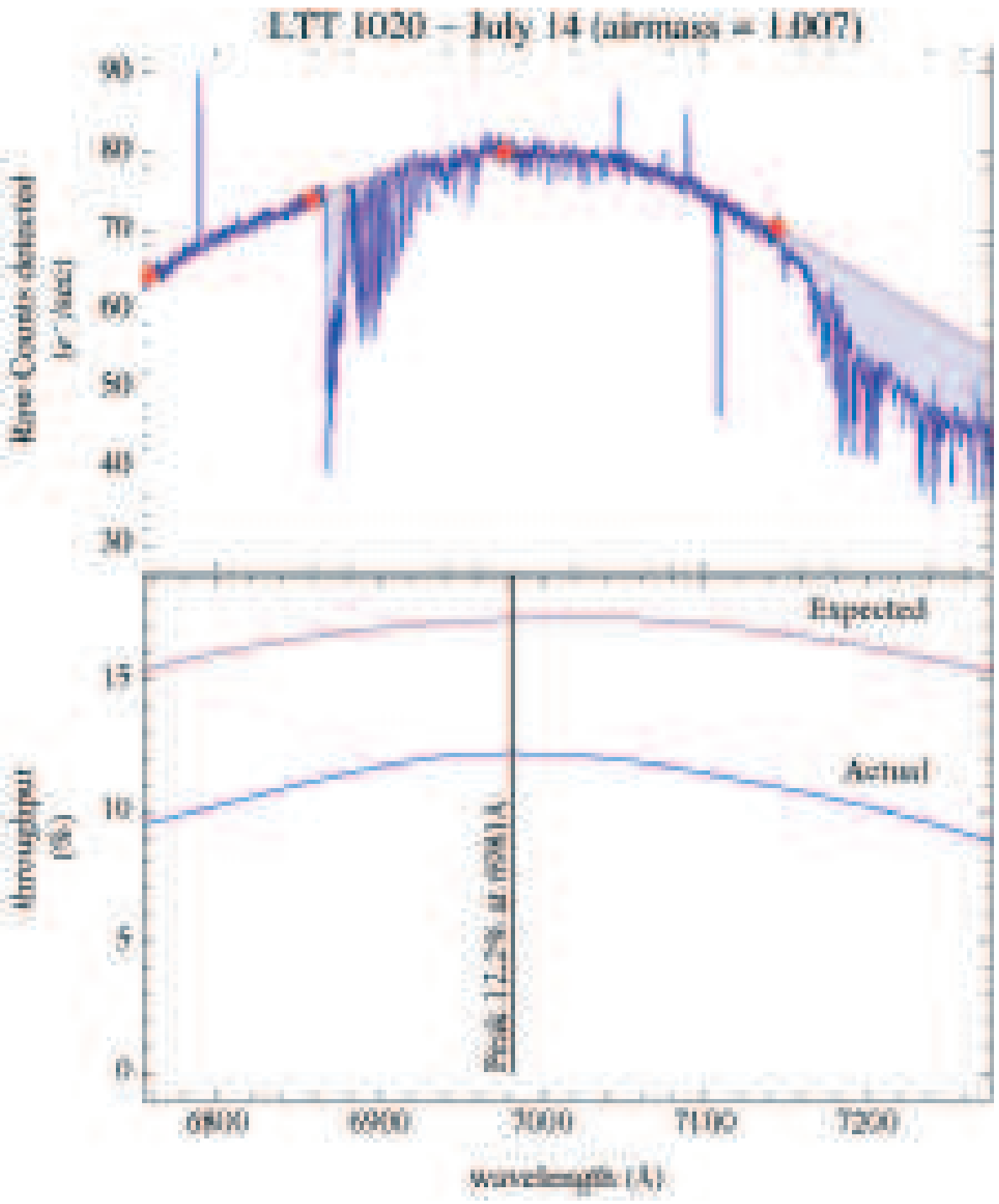

Figure 2: One of our standard star observations demonstrating SPIRAL's throughput. The first plot shows the raw e⁻ counts/sec from the detector (converted to electrons and divided by the total exposure time), and the smoothed continuum level defined by the marked points. Note the numerous atmospheric absorption lines in this wavelength range which we interpolate over (shading). The second plot shows the measured throughput of the whole system vs wavelength, and the expected throughput based on initial estimates presented in the online exposure time calculator. This star (LTT 1020) was observed near zenith.

# PHOTONIC OH SUPPRESSION OF THE INFRARED NIGHT SKY: FIRST ON-SKY RESULTS

J. Bland-Hawthorn, S. Ellis (U. Sydney),
R. Haynes, A. Horton (AAO)

## The problem and a solution

Imagine a world in which the sky is always bright, a world in which there is no night. In such a world, we would see very little beyond the glowing atmosphere. On Earth, for example, we would only see the Sun and the Moon, a few planets and maybe a few stars. Everything else would be invisible to us and we would know little about the distant universe. Of course, this ***is*** the world we live in – at infrared wavelengths. The infrared sky is bright, day and night. This is a major obstacle to astronomers because a great deal of information about the most distant universe emerges in this part of the spectrum. One approach to bypassing the atmosphere is to launch a space telescope to get above it. This option, however, is hugely expensive and the size of an orbiting telescope is limited.

A solution to the apparently insoluble problem of the infrared night sky emerged in 2004, when our team from the Anglo-Australian Observatory[1] began to explore new developments in photonics – the science of manipulating light in materials like optical fibres – which lies at the heart of modern telecommunications (Bland-Hawthorn et al. 2004; Opt Exp 12, 5902).

The important step is to realize that the infrared sky is bright because the atmosphere glows in hundreds of very narrow spectral lines. Without these lines, the night sky would appear very dark (Ellis & Bland-Hawthorn 2008; MNRAS, 386, 47). So how do we suppress such a large number of night-sky lines by factors of hundreds to thousands in an efficient way?

The trick is to focus light from the telescope onto an optical fibre. The light then bounces down the fibre before being split up into parallel tracks. Along each track, the light is processed by a fibre Bragg grating. This extremely efficient device lets the "good" light from between the atmospheric emission lines go onwards, but reflects the "bad" light from the emission lines back in the direction whence it came. The parallel tracks holding the "good" light join up again into an optical fibre that can then be fed into an astronomical instrument. We call this remarkable device a "photonic lantern" (Leon-Saval et al. 2005; Opt Lett 30, 2545): this breakthrough technology allows a single-mode action in a multimode fibre. The way it works is described in an earlier AAO newsletter – http://adsabs.harvard.edu/abs/2005AAONw.108....4B .

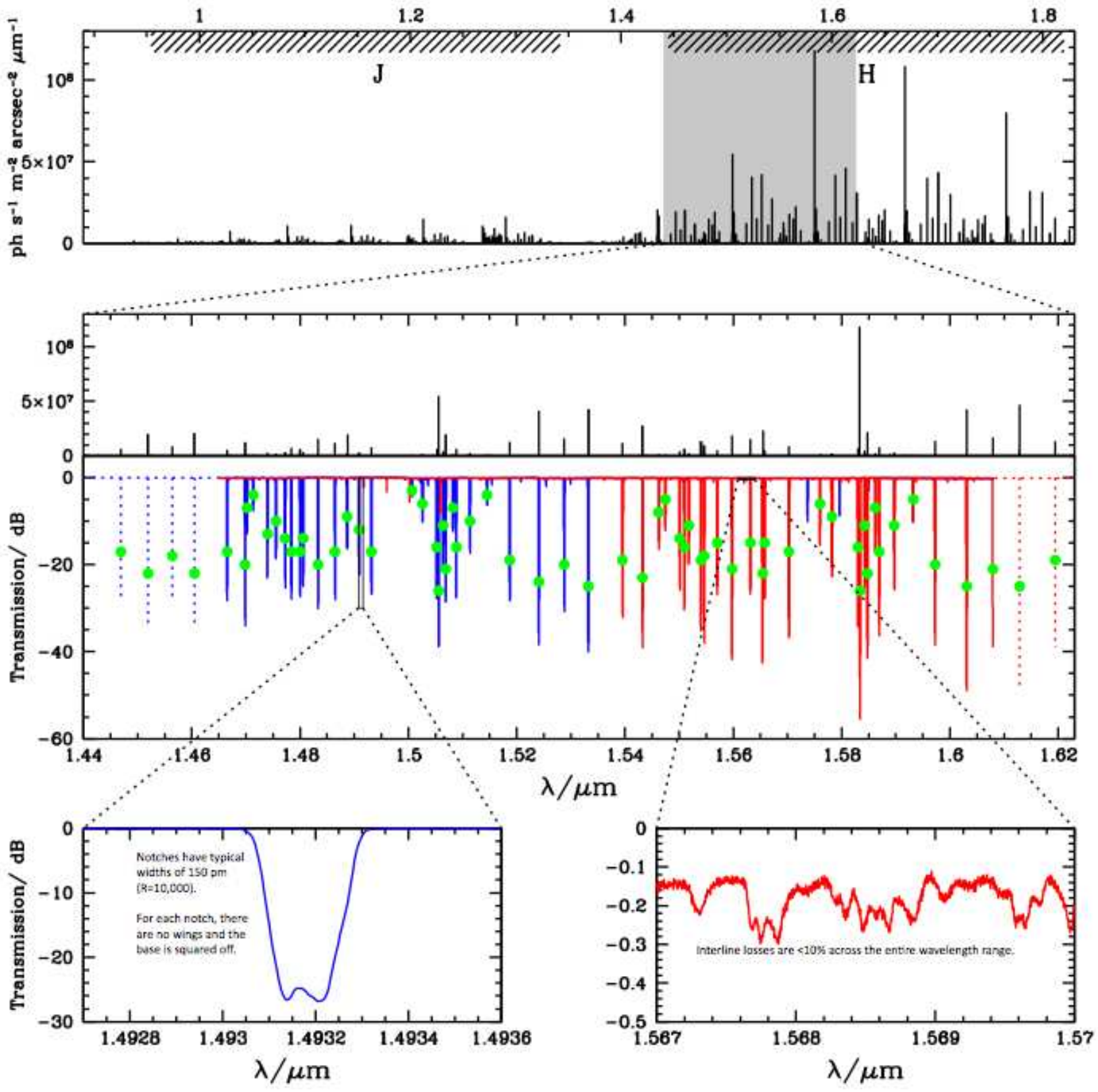

Figure 1: The H-band region targeted for OH suppression is shown in grey (top panel). This region is expanded in the upper middle panel to allow an easy comparison with the grating design in the lower middle panel. The grating is "overcooked" compared to the optimum notch strengths shown by the green dots. The bottom left inset illustrates how each notch is roughly a square profile; the bottom right inset shows how the interline loss is less than 10%.

[1] Two members of the original team have since moved to the University of Sydney.

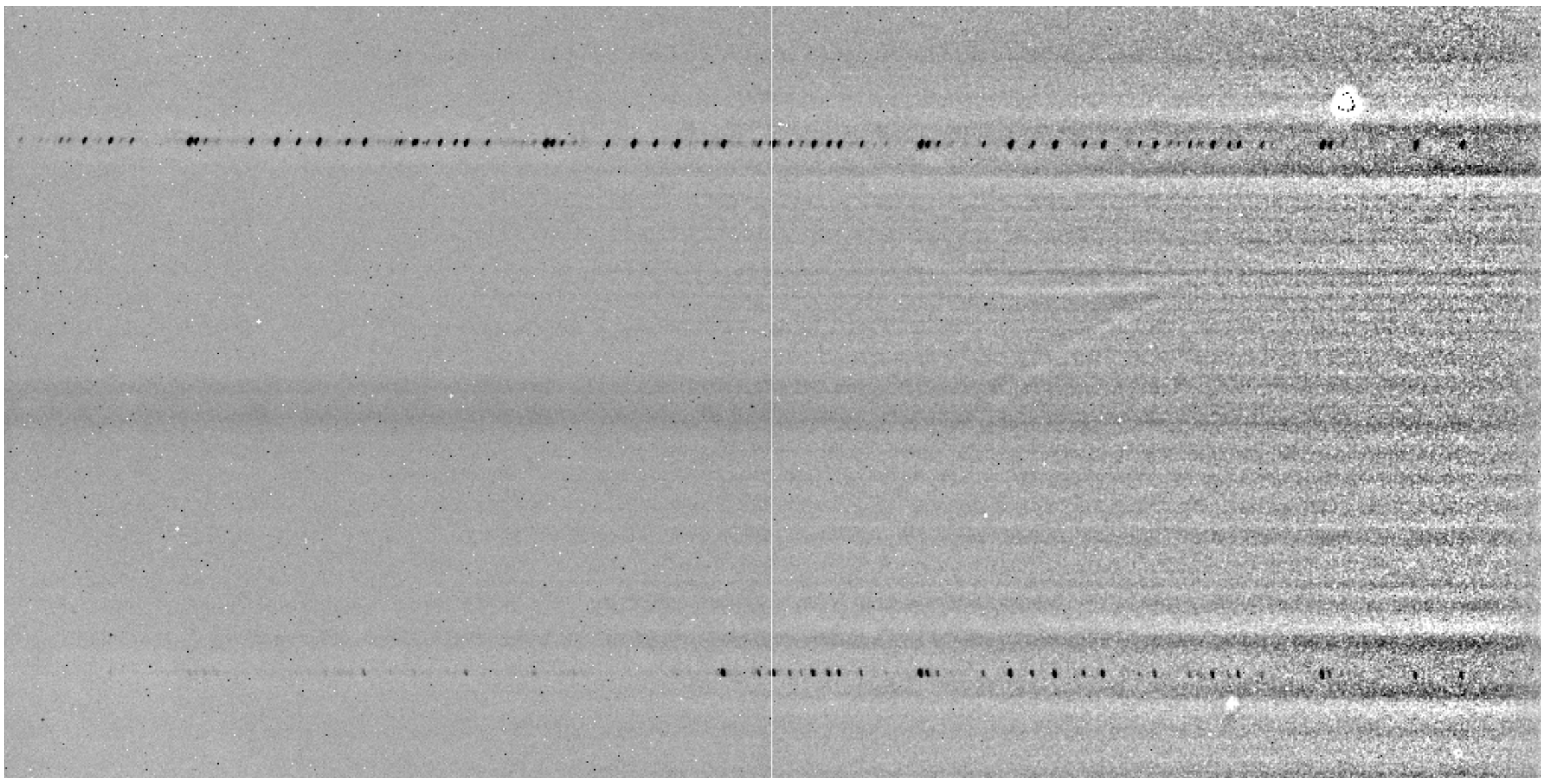

Figure 2: The stacked raw IRIS2 image showing the unsuppressed night sky spectrum (top strip of dots) and suppressed sky (bottom). The wavelength direction is horizontal (wavelength increasing to the right) and covers almost the full astronomical H-band at near-infrared wavelengths. The lack of dots in the left half of the bottom spectrum is a wonderful demonstration of photonic OH suppression; the region to the bottom right was not suppressed. Some of the suppressed region falls off the detector to the left.

The transmission profile of the fibre Bragg grating is shown in Figure 1. The top panel shades the region of the spectroscopic H-band that we attempt to clean up with the OH suppression gratings. The middle panel shows how the expanded spectrum compares with the grating notches: these "reflectors" line up (almost) perfectly with the OH sky lines. Note that we are matching the notch strengths to the OH line intensity, i.e. strong notches for strong lines, weaker notches for weaker lines. In the course of our work, we discovered that this minimises the amount of "information" in the grating design which helps us to maximise fibre throughput and manufacturing yields. The green dots indicate the ideal notch strengths. Unfortunately, our gratings exceed the target strengths by a factor of 10 which causes some of the notches to drift a little in wavelength and also increases the insertion loss within the gratings. This is easily remedied in the next generation of gratings.

**Do OH-suppressing fibres actually work?**

No amount of lab testing can really replace an on-sky experiment. In order to find out, we conducted an experiment on the 4th floor of the AAT in the week before Christmas (Dec 17–19, 2008). To assist us, the site staff drilled a hole through the dome structure to allow a direct feed from outside to the IRIS2 infrared spectrograph. This is the first time that the dome structure has been breached (tantamount to drilling a hole through a ship's superstructure) and it is rumoured to have been a traumatic experience for the drillmaster.

The OH-suppressing fibre, and a comparison fibre without a suppressor but with a photonic lantern in series, were fed through the dome hole and pointed at the sky. The light from the other ends of the fibres was re-imaged onto the focal plane of IRIS2. The spectral region of interest was isolated with an H-band spectroscopic filter (1475–1800nm). The fibre core size is 60 microns (numerical aperture NA=0.1) and the photonic lanterns use 7 parallel single-mode tracks holding identical gratings. The fibre Bragg gratings knock out 63 OH lines at a resolving power of 10,000. The IRIS2 spectral resolution was set at R=2500.

*And the resulting spectrum is a fabulous confirmation of the technology!* Even after a 15 min exposure, it was clear that the OH-suppressing fibre works. A stacked IRIS2 image is shown in Figure 2. Unfortunately, we could not get all of the suppressed region onto the detector. Our experiment effectively cleaned up half of the spectroscopic H-band at low loss (~10%). A test of a complete H-band (and for that matter J-band) OH-suppressing fibre will be carried out later in the year.

**Results**

The spectrum with and without suppression is shown on the front cover. Note several important points:

1. The OH lines are being suppressed at a resolution that is 4x higher than is seen here, i.e. R=10,000. Some of the suppressed region is missing to the blue (left).

2. Note that there is *no* noise residual, a remarkable testament to photonic OH suppression.

3. Since the fibre is being pointed at the sky, the weak continuum in the suppressed spectrum is the infrared starlight falling within the 10-degree fibre beam. We are therefore unable to measure the OH inter-line continuum with this experiment.

4. The slight mismatch in suppressing 2 or 3 of the OH emission lines is due to the fact that the fibre Bragg gratings were "overcooked," which causes a few of the notches to saturate and drift slightly. This saturation also partly accounts for the some of the loss between the upper and lower spectrum in Figure 2, something that is easily corrected in future gratings.

**Where next?**

The theory of ultra-broadband grating design is highly complex and is discussed elsewhere (Bland-Hawthorn et al. 2008; JOSA, 25, 153): *our gratings are by far the most complex optical filters ever constructed in any field of science,* as we found out when investigating the prospect of international patents. Each suppressing notch has a squared-off profile with no detectable "ringing" effect (see Figure 1, inset). The Bragg gratings are manufactured by Redfern Optical Components (ROC) in Sydney to our prescription.

The team is now working with ROC to perfect and mass-produce the new technology, and to install it on as many of the big telescopes as possible. We hope to start with the Gemini 8m telescopes in Hawaii and Chile operated by the Gemini consortium, of which Australia is a member. First light with a science fibre is expected to take place in 2010 and will suppress up to 400 OH lines over the J+H spectroscopic bands. The fibre will be fed to an existing IR spectrograph as outlined in Horton et al. (2008; SPIE, 7014, 104).

**Science impact**

The new sky-suppressing fibres will render the sky at least 30 times darker; the actual improvement will not be known until the inter-line continuum is measured accurately next year. A detailed discussion of the scientific implications is presented by Ellis & Bland-Hawthorn (2008). We were invited to contribute to a section of the US Decadal Review on Astronomy through the GSMT Science Working Group: this article can be accessed at http://www.gsmt.noao.edu/swgt-suc.php .

**Astrophotonics**

The use of ultra-broadband fibre Bragg gratings and photonic lanterns are two examples of the new field of astrophotonics that lies at the interface of astronomy and photonics. This burgeoning field is the focus of the Feb. 2 issue of Optics Express (vol. 17:3) and is reviewed in the editorial article by Bland-Hawthorn & Kern (2009): all articles are free to air at http://www.opticsinfobase.org/oe/Issue.cfm . Another astrophotonic device is the imaging fibre – or hexabundle – that will be featured in the next focus issue on astrophotonics later in the year. The first on-sky tests were carried out after the OH suppression tests and these results will be presented in the next issue of the AAO newsletter.

**Funding** This work is supported by a Federation Fellowship from the Australian Research Council and a grant from the Science & Technology Facilities Council (UK).

## AUSGO CORNER

Stuart Ryder (Australian Gemini Office, AAO)

**Semester 2009A**

In this semester we received a total of 24 Gemini proposals, of which 15 were for time on Gemini North or exchange time on Keck or Subaru, and 9 were for time on Gemini South. The oversubscription factors were 2.25 in the north, and 1.70 in the south. At ITAC, Australia was able to schedule one classical run each on Subaru and Keck, with 16 more programs going into Bands 1–3.

For Magellan we received 10 proposals for a very healthy oversubscription of 2.9. The recommendation by AAL's Astronomy NCRIS Strategic Options Committee (ANSOC) in September 2008 that AAL commit US$1.72 million to continuing Australia's access to Magellan through until the end of 2011 was a welcome endorsement of the scientific and strategic value that this access brings.

Some incorrect information about the instrumentation available on Magellan in 2009A (and in particular a late decision by the Magellan Council to withdraw LDSS3 with immediate effect) necessitated a one week extension to the proposal deadline. At its latest meeting, ATAC agreed there was merit in having the Magellan proposal deadline one week later than the Gemini proposal deadline, so this will become permanent from Semester 2009B onwards.

**Staffing**

Dr Terry Bridges has resigned as Deputy Gemini Scientist at the AAO and returned to his native Canada for personal reasons. We are very sorry to lose Terry's expertise and enthusiasm so soon after his arrival, but wish him well in his future endeavours. Terry was instrumental in getting the Australian GMOS Imaging Contest (see below) off the ground, and we are grateful that he has agreed to stay involved in this project.

We have nevertheless been fortunate to secure the services of Dr Simon O'Toole at short notice as Terry's replacement. Simon has been the Anglo-Australian Planet Search Research Fellow based at the AAO since late 2005 funded by an STFC grant to Chris Tinney (now at UNSW) and Hugh Jones (U. Hertfordshire). Simon was rapidly brought up to speed on all the various modes of GMOS by Terry before his departure, and his background in optical spectroscopy of subdwarf stars and precision radial velocities will make him a real asset to AusGO.

**GMOS Imaging Contest**

As its contribution to the International Year of Astronomy, AusGO will be running a contest for Australian high schools to win one hour of time on Gemini South. Modelled on the hugely successful Canadian Gemini Office contests that have been run in recent years, our contest will invite Australian secondary school children to nominate one object they would like to observe with GMOS, and explain why on scientific merit and aesthetic grounds. A Semester 2009A proposal submitted to ATAC by Terry Bridges on behalf of AusGO for the required observing time in winter 2009 was the top-ranked program in Band 1, and has been awarded rollover status. We are grateful to ATAC for recognizing the high visibility and impact that a full-colour Gemini image involving school kids can have in raising the profile of astronomy, and Gemini in particular, within Australia in the IYA. Details on how to enter and the judging process are still being worked out, but will appear on the AusGO web site early in the new school year.

**AGUSS**

For the third successive year AusGO has offered talented undergraduate students enrolled at an Australian university the opportunity to spend 10 weeks over summer working at the Gemini South observatory in Chile on a research project with Gemini staff. The Australian Gemini Undergraduate Summer Student (AGUSS) program is generously sponsored by AAL. After promoting AGUSS on the ASA e-mail exploder as well as a new poster circulated to Australian physics departments, we received a total of 20 applicants, of whom only two could be selected. The 2008/09 AGUSS recipients are Sophie Underwood from the University of Adelaide, and David Palamara from Monash University. They are due to present the results of their research via video link from Chile to Australia in mid-February 2009.

**Magellan Fellowships**

The two current Magellan Fellows, David Floyd and Ricardo Covarrubias, are due to complete their Magellan support duties in Chile later this year. David will spend his final year carrying out full-time research at the University of Melbourne, while Ricardo will be working at the AAO. As part of the extension of Australian access to Magellan mentioned above, there will be a second round of Magellan Fellowships, involving up to 2.5 years in Chile working at Magellan, followed by a final year of research at an Australian institution of their choice. Applications for Magellan Fellowships close on 16 February – see http://www.aao.gov.au/jobs/magellanfellow-nov08.html for the full advertisement.

**Next Gemini Science Meeting**

The Subaru and Gemini Observatories will be hosting a Joint Subaru/Gemini Science Conference in Kyoto, Japan from 18 – 21 May 2009. This will be followed by a Gemini Users meeting on 22 May. The aims of this conference are to promote mutual understanding between the user communities, awareness of each observatory's facilities and scientific strengths, and foster new international collaborations. The meeting also comes at a crucial time for the WFMOS instrument concept, which if approved would be funded jointly by Gemini and Japan, and installed on Subaru. AusGO intends to offer some travel support to Australian participants to ensure Australian Gemini and Subaru science is well-represented at this forum. Anyone interested in attending is urged to register at http://www.kusastro.kyoto-u.ac.jp/kyoto2009/info.html as early as possible.

## THE STATE OF THE AATAC

Karl Glazebrook, Chair AATAC

The Anglo-Australian Time Assignment Committee is charged by the AAT Board with allocating the majority of the time (90%) on the AAT. This committee of 5 Australian and 2 UK astronomers work together to rank proposals and allocate the Australian and (rapidly diminishing) UK shares of the telescope.

The landscape for AAT science at the beginning of 2009 is remarkably healthy, the demand for AAT time remaining heavy with typical oversubscription factors per semester[1] at around 2x. Given the heavy use of the telescope by Large Programs the *marginal oversubscription* can be even higher approaching 3x. On a more qualitative note I see every meeting that the level of science being proposed for the AAT is quite diverse and extraordinary. A typical comment by a new AATAC member is along the lines of '*I didn't know how much good science was still being done on the AAT.*'

The Large Programs themselves are a statement of this success. We are currently running two of these, WiggleZ and GAMA; both are extra-galactic but with quite different goals. WiggleZ is reaching back 6 billion years to measure the fundamental cosmology of the Universe, while GAMA is aiming to measure the roles of baryons and dark matter in assembling today's Universe. It is certainly possible we will entertain future Large Programs, but given the big time demand (almost a quarter of the total telescope time) from WiggleZ and GAMA, these would almost certainly have to be bright time/stellar projects to have a chance of even being schedulable! WiggleZ and GAMA themselves are now scheduled to finish in 2010A (noting that GAMA is only an A semester project).

Given this huge demand for AAT time how can your proposal stand a chance of success? There are several factors to be aware of when writing proposals which can help increase your chance of success.

Firstly, write to your audience. AATAC members have research fields ranging from cosmology to planets. Every semester we have 30–40 proposals to read and no more than 5–10 minutes will be spent on each proposal. Everyone grades all the proposals they read and the average grade determines the initial ranking. The introduction and big picture should speak to a general astronomical audience with enough subsequent technical detail to satisfy the experts on the committee that you know what you are talking about!

Avoid jargon and especially acronyms. Cosmology proposals talking of BAOs, HODs, P(k)s, JDEMS, LSSTs and DESs are one notorious example. Others are stellar proposals which talk about things like XY Boo stars without explaining what they are physically, or their scientific importance.

Try to frame your proposal in terms of a specific *scientific experiment* with a *defined outcome*. This is good: "This proposal will measure the bogosity of XY Boo stars to 5%, a factor of ten better than currently determined, providing a clear distinction between the Snark and the Boojum formation model". This is bad: "By obtaining a bigger sample of XY Boo stars we will better constrain their formation mechanism." Words like "constrain" suggest a lack of definiteness, both in terms of immediate goals and how much telescope time is really needed. Avoid the classic mistake of looking like you are going on a '*fishing expedition*' (unless of course you are after cosmic coelacanths in which case we might be convinced). In particular if you can justify numbers of objects or telescope time quantitatively (e.g. via a simulation) this will really solidify the proposal.

If you have had AAT time in the past for anything related, show clear evidence that you have spent time working on the data and used this information to frame the present proposal. AATAC is very forgiving of cloudy nights and instrument failures, but not of laziness or a lack of postdoc students!. We are a sophisticated audience and have similar or worse issues in our own careers.

Pay attention to your feedback if you have been unsuccessful with a proposal for a particular project in the past. This is the first thing we check. I will admit sometimes AATAC feedback is not all that helpful, writing useful feedback is surprisingly difficult and sometimes all we can say is 'great science, but not ranked high enough'. It is often a mistake to over-analyse feedback and think "all I have to do is fix item xxx and then I will get the time!". That's not how it works. But if there are particular scientific or technical problems these must be addressed, and it always helps to really present the greater scientific importance.

[1] Defined as the ratio of available nights to requested nights. If one adopts the ESO definition and uses clear nights (averaged) then the numbers are 1.33x higher.

Finally if you are after dark or grey time in Feb/Mar/April please bear in mind that there are at most 4–5 nights left in this busy period after WiggleZ and GAMA are taken off the top. Unless your proposal is the very top it won't succeed. There is a similar, but somewhat easier competition for the Aug/Sep/Oct dark/grey time due to the absence of GAMA, .

AATAC is continually evolving as the UK share has been continually diminishing since 2005 and comes to an end in 2010. The formulae we have used for allocating time and charging time to partners have also evolved to cope with this. They attempt to balance the competing demands of fair shares for Australian and UK users, the desire to maintain a fraction of 'open skies' and the desire not to disadvantage UK users against international applicants. It pays to read the AATAC guidelines (which are set by the AAT Board) in detail to see how this works.

Let me conclude by saying that AATAC is always looking for new members. There will be a couple of vacancies in the next year or two and one of the best ways to learn how to write really good proposals is to serve on a committee like this and see – well – all the mistakes people make. If you are interested then look out for calls for nominations to AATAC on the ASA and STFC mailing lists.

## LARGE OBSERVING PROGRAMS ON THE AAT

**Request for Proposals Semester 09B**

A new *Request for Proposals* is being issued for major new observing programs to commence in semester 09B (August 2009 to January 2010) or semester 10A (February 2010 to July 2010). Programs extending beyond semester 10A may possibly be subject to modification following the end of the current AAT Agreement on 30 June 2010.

The AAT Board (AATB) encourages ambitious large programs and does not set an upper limit on the fraction of time large programs can be awarded. The AATB expects large observing programs to be awarded *at least* 25% of the available time on the AAT in semesters 09B and 10A; this fraction of time will be reviewed for subsequent semesters.

Proposers are encouraged to form broad collaborations across the Australian and British communities in support of their programs; other international collaborators are welcomed. The status of current large observing programs can be found at http://www.aao.gov.au/AAO/astro/apply/longterm.html.

All proposals will be evaluated by the Anglo-Australian Time Allocation Committee (AATAC), and should use the standard AATAC form (although with non-standard page limits). AATAC will award time based on considerations including the relative scientific merit and impact of the large programs and standard programs, the quality of the management, publication and outreach plans, and the phasing of programs to provide a steady rollover of large programs for the longer term.

**Proposals for large observing programs should be submitted to AATAC by the standard proposal deadline of 15 March 2009.**

**Proposers should consult the *Request for Proposals* for more information about the process: http://www.aao.gov.au/AAO/astro/Large_Programs_RfP_09B.pdf.**

**Anyone considering submitting a large program proposal should contact the AAO Director (director@aao.gov.au) in advance to discuss their plans.**

# NEW ARRIVALS AT THE AAO

There have been a number of new arrivals at the AAO since the last newsletter. The Epping News column on page 23 provides some information, but Andrew Hopkins and Chris Springob introduce themselves here.

**Andrew Hopkins – Head of AAT Science**

In November 2008 I joined the AAO as the new Head of Anglo-Australian Telescope Science. I grew up in country New South Wales, and completed my undergraduate science degree and PhD at the University of Sydney. I took up a postdoctoral position at the University of Pittsburgh in 1999, during which I was awarded a Hubble Fellowship by the Space Telescope Science Institute. After spending six years living in Pittsburgh, which it turns out is actually a rather pleasant city contrary to my preconceptions, I returned to the University of Sydney in 2005 to take up an ARC Queen Elizabeth II Fellowship in the School of Physics. The Head of AAT Science position gives me the opportunity to build on my experience with the AAT and to work with the astronomy group at the AAO.

My scientific interests focus on galaxy evolution research, which I approach from a broad multiwavelength perspective. I have a particular interest in understanding how and when stars in galaxies formed, and how this depends on the masses and environments of the galaxies. An open question here is where the gas that fuels the star formation comes from. One of the interesting puzzles that I have recently been exploring is whether a burst of star formation always forms stars with the same distribution of masses or not. In particular, are bursts of star formation in the distant past similar to those that happen today? This long-held assumption is now starting to be questioned. To answer these questions I am involved in several international observational survey collaborations, including the GAMA (Galaxy And Mass Assembly) project, featured in the previous AAO Newsletter. I also work on various radio survey projects, and collaborate closely with many ATNF researchers, a valuable opportunity to combine the strengths of the AAO with that of ASKAP, the next-generation radio telescope facility being built by ATNF in Western Australia. The strength of multiwavelength data is well-known, in particular the combination of radio and optical survey programs, and this underpins the scientific motivation of the major proposed ASKAP surveys.

In addition to my research, I have a passion for communicating science to school students and the general public. In 2006 I was awarded a Young Tall Poppy of the Year award by the Australian Institute of Policy and Science, giving me the opportunity to present my research to secondary schools around the state. I am also an active participant in several other science outreach programs, including Scientists in Schools and MyScience, all aimed at increasing awareness of and interest in science in primary and secondary students.

I am excited to be joining the AAO at a time of new opportunities, including the development of the new HERMES instrument and surveys like GAMA. I am enjoying working in what I find to be a particularly inspiring and dynamic environment, and I look forward to supporting and developing my research in conjunction with the talented and accomplished team at the AAO.

**Chris Springob – AAO Director's Fellow**

In September of 2008, I moved to Australia to begin a 3 year fellowship at the AAO as the Director's Fellow. I completed my PhD at Cornell University in 2005, and spent the following three years in postdoctoral positions at the US Naval Research Laboratory and Washington State University. I've spent most of my career to date working on redshift-independent distance indicators to galaxies, particularly the Tully-Fisher relation. I was hired to work at the AAO on the 6dFGS Fundamental Plane Survey, which will be the largest survey of galaxy peculiar velocities ever compiled. The survey will help us to resolve longstanding questions about the large scale structure of the universe, and how large scale motions in the universe can be reconciled with the $\Lambda$CDM standard model.

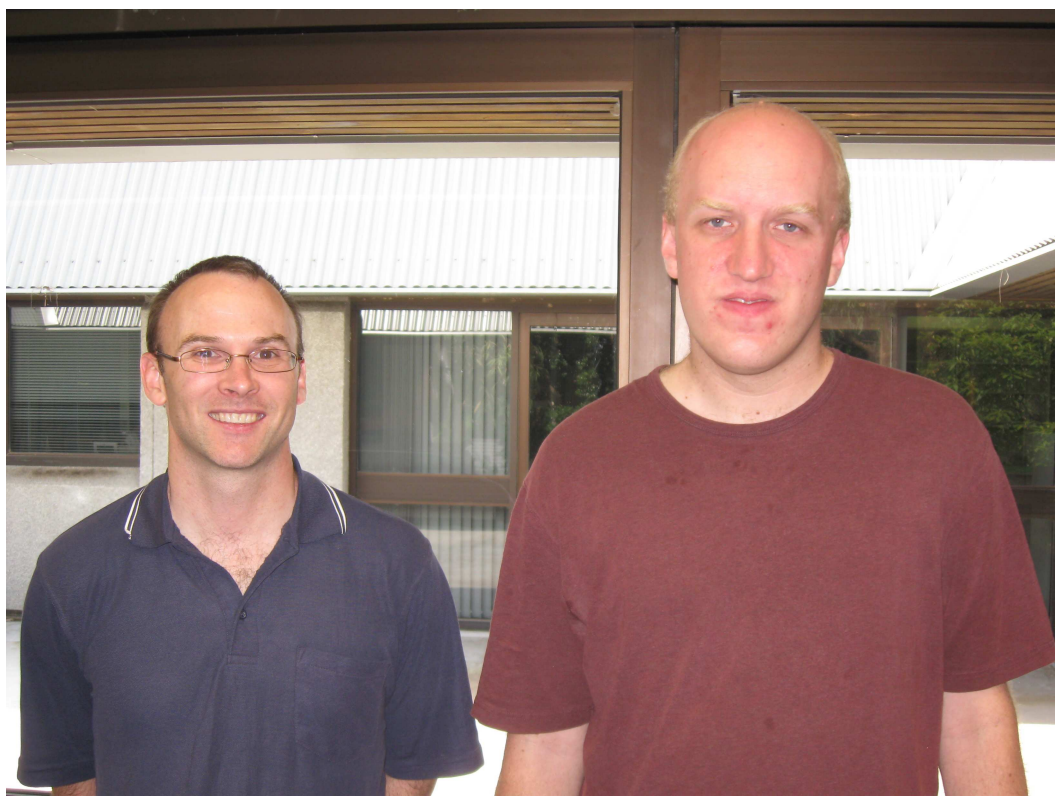

Andrew Hopkins (left) and Chris Springob

## SUMMER STUDENTS

Paul Dobbie

The AAO runs a twice yearly fellowship programme to enable undergraduate students to gain 10–12 weeks of first hand experience of astronomical related research.

Chris Owen is a third year physics/astrophysics student from the Queensland University of Technology. He is working on reducing photometric and spectroscopic data of the unusual Type IIn supernova 1978K under the supervision of Stuart Ryder. Regular monitoring of the supernova's radio flux between 1991 and 2008 has revealed an increasing level of deviation from a smooth decline in recent years. This indicates a change in the mass-loss rate of the progenitor star – most likely a Luminous Blue Variable – prior to its demise. The outcome of this work will determine whether there is a corresponding change in the supernova's optical spectrum. During his time at the AAO, Chris visited the AAT and the ATCA where he had the chance to see the telescopes and observers at work.

Madusha Gunawardhana is a Macquarie University undergraduate, who is due to start her honours year in Astronomy/Astrophysics at the beginning of March. Under the supervision of Andrew Hopkins, she is using the H-alpha line luminosity corrected for stellar absorption and obscuration to constrain the star formation rate in a large sample of galaxies drawn from the Galaxy And Mass Assembly year one observations and the Sloan Digital Sky Survey. The ultimate goals of her project are to explore the SFR density of the universe as a function of galaxy mass, over a modest range in redshift, corresponding to about 4 Gyr, or roughly a third of the age of the universe. Additionally, she hopes to improve understanding of the impact of the lowest mass galaxies in the production of stars in the local universe and their influence on the mass dependence of the SFH up until the present epoch.

Allar Saviauk, who is originally from Finland, is a mechanical engineering student of astronomy and optics at Macquarie University. The aim of his project is to investigate the causes of and identify potential solutions to the fringing which occurs in a proportion of AAOmega spectra. His work, with Roger Haynes, has confirmed that during periods of high temperature the bonds between some fibres, which carry light to the spectrograph and the prisms which direct light from the target down the fibres, break, resulting in air gaps between these two system components. A number of new designs have been proposed and Allar has travelled to Siding Spring on a number of occasions to test these. He hopes to be involved in the design and development of telescopes and instrumentation in the future.

Liliana I. Lopez has recently received her masters degree from San Francisco State University and completed a short research project at the Institute of Astronomy, University of Tokyo. She has been working on the development and testing of an algorithm to resolve the prevalent issue of atmospheric dispersion in the spectral images observed with the Gemini-GMOS instrument in the IFU mode. During the testing phase of the project her analysis of two quasar targets (one with each of GMOS-N and GMOS-S to assess instrumental effects) has identified peculiar emission line features that reveal "line-splitting", which appears to indicate possible ionization cones and outflow structures in the host galaxies. At the end of her project, Liliana will return to California to start a PhD program.

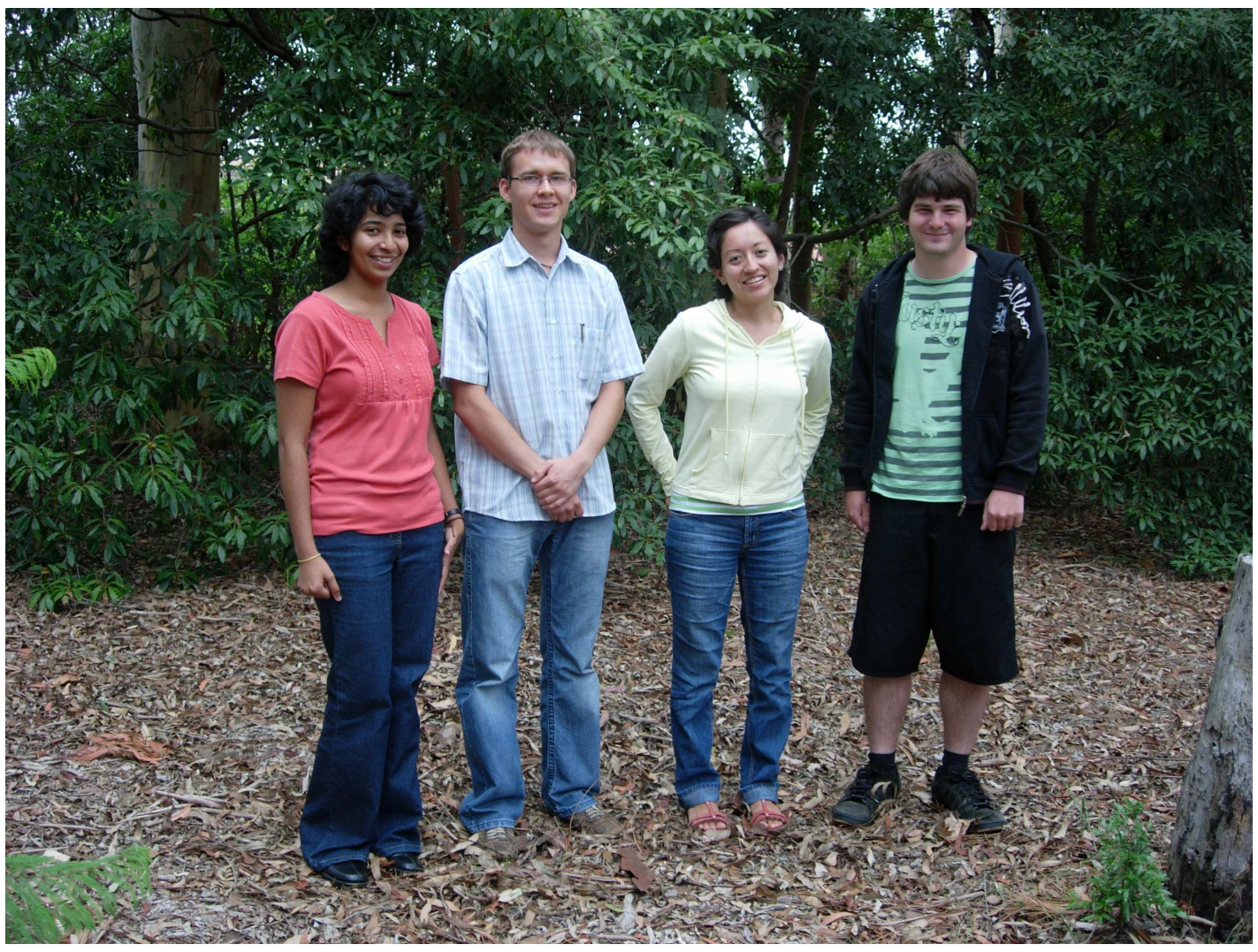

Summer students L to R: Madusha Gunawardhana, Allar Saviauk, Liliana I. Lopez and Chris Owen

## LETTER FROM COONABARABRAN

Rhonda Martin

Variable weather this summer has made for many a disgruntled observer although the resulting sound and light shows of some rather awesome storms have made up for disappointment – partly, anyway. But finally the heat has arrived with clear nights and whoops of joy have been heard coming from the control room as the data rolls in. There was even an annular eclipse to be seen last week – Steve Lee took a great shot of it too.

Our new people – Guy Andrews, Darren Mathews, Steve Chapman and John Goodyear – have all settled in well and are welcome and excellent additions to our staff.

The Refurbishment project is moving along in leaps and bounds resulting in a much more ergonomic workplace. The library on the 1st Floor has now been turned into a state-of-the-art conference room and is a much pleasanter place than the old room on the 2 ½ Floor which used to send people to sleep because of its stuffiness. The library is now situated here and oddly enough, they fit together and the floor looks like a proper library with a lot more space than the old one.

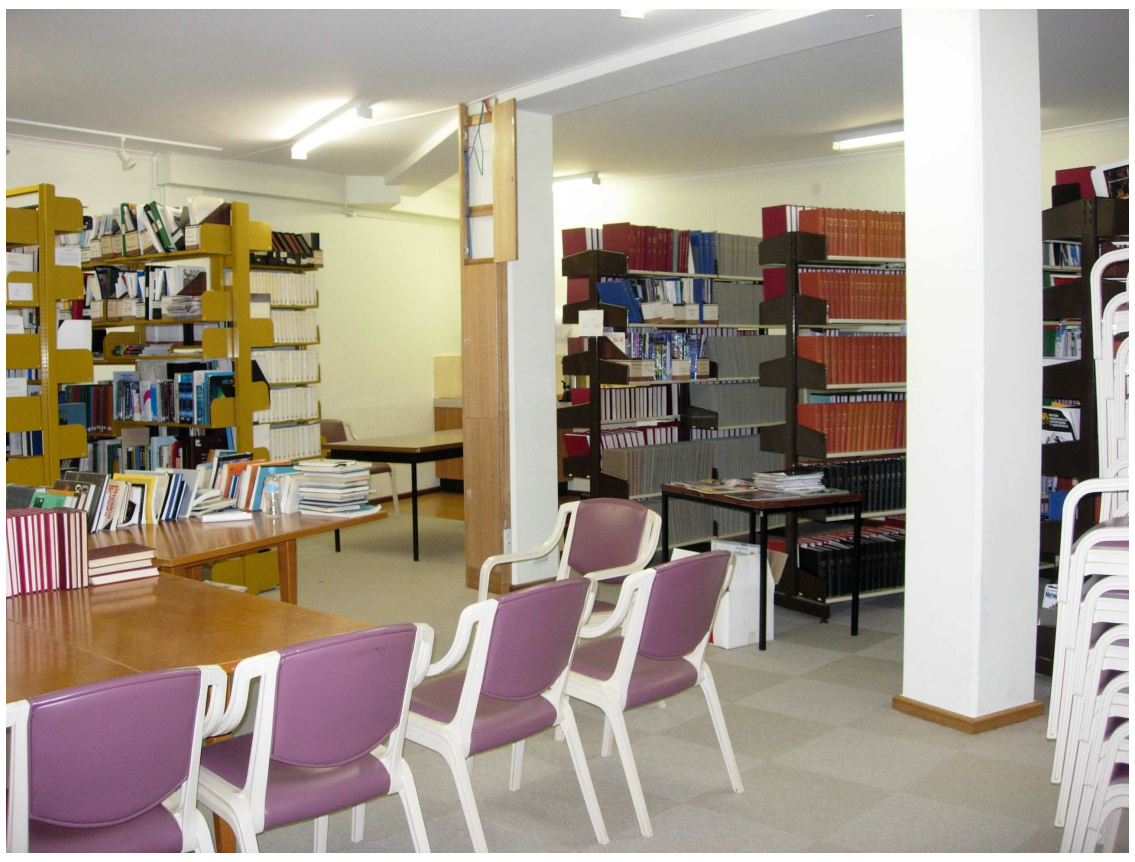

Old conference room becomes new library at the AAT.

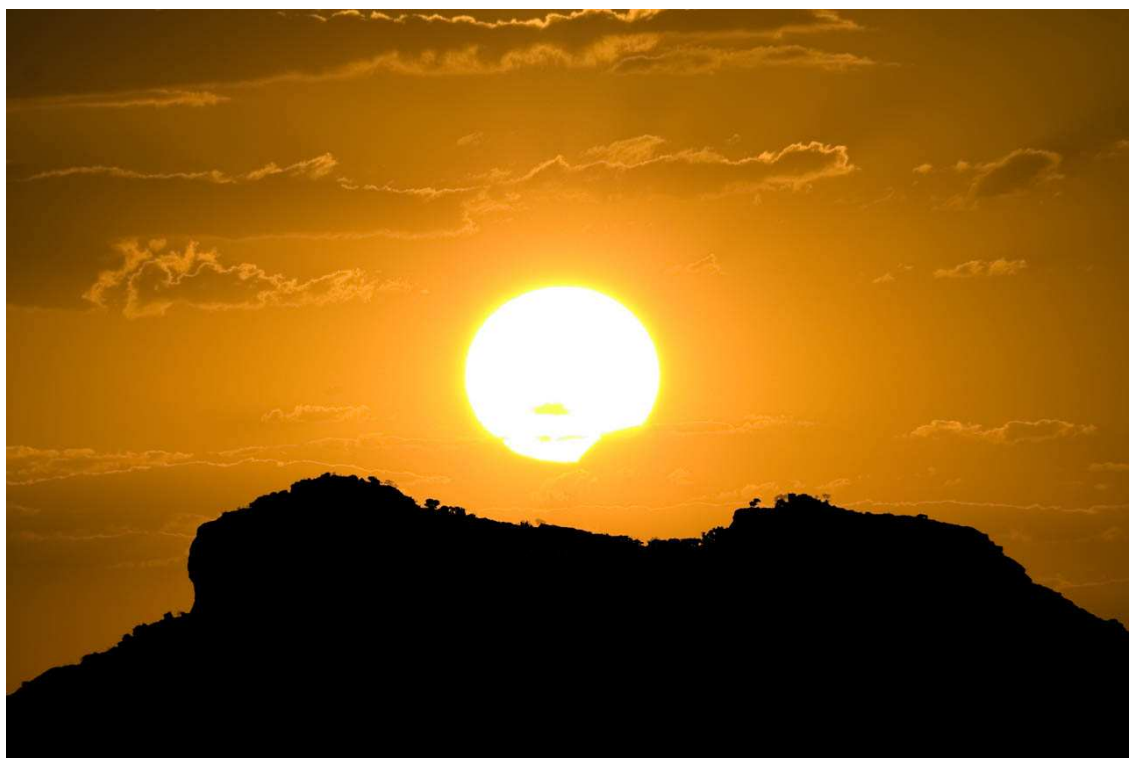

Annular eclipse. Photo courtesy of Steve Lee.

## EPPING NEWS

Sandra Ricketts

Once again, this column has largely become a list of arrivals and departures.

Andy Bunker returned to the UK in September 08 and our new Head of AAT Science is Andrew Hopkins, who introduces himself on page 21.

The new AAO Director's Fellow, Chris Springob, replaces Heath Jones, now a Research Astronomer, as mentioned in the last newsletter. Chris also introduces himself on page 21.

Jon Lawrence has joined the AAO instrument science group in a joint position with Macquarie University, having spent the last eight years at the University of New South Wales developing instrumentation and power and control platforms for remote astronomical site testing on the Antarctic plateau. Jon also spent time last year at the AAO due to his role as project scientist for the PILOT telescope design study.

And we will greatly miss Will Saunders although he threatens to reappear from time to time. As Will says, there have been many memorable times, both good and bad!

Sue-Ellen Fahey left at the end of the year, and we welcome two part-timers – Cathy Cafe, who has the very important task of looking after the payroll, and Suzanne Tritton, who will be our Personnel Officer.

Andrew McGrath leaves us after seven years for Flinders University in Adelaide, returning to his roots in meteorology. An enjoyable lunch was held at the local pizza restaurant to farewell Andrew.

An era at the AAT ended last month when it was announced that the Interdata computer, replaced by the new TCS, was to go to the Powerhouse Museum in Sydney. A pleasing and fitting destiny!

And Fred Watson has added another award to his collection – the Queensland Premier's Literary Award in the “Science Writer” category for his book “Why is Uranus Upside Down?” Congratulations, Fred!

Finally, just before Christmas, the AAO staff at Epping gathered for lunch at a local restaurant to celebrate the holiday season.

THE BACK PAGE

# The heart of the Milky Way

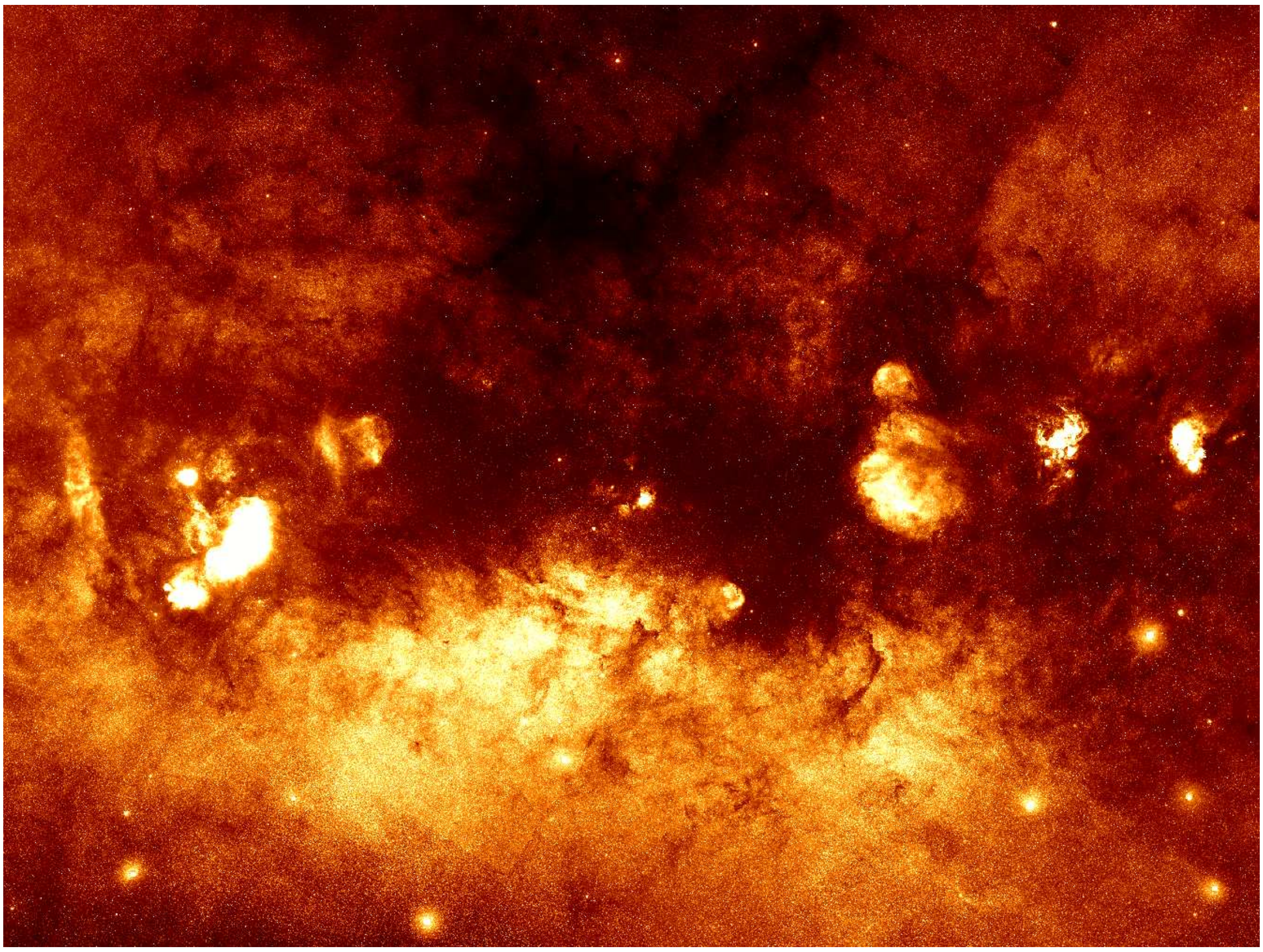

The central 20 by 15 degrees of the Milky Way viewed by the AAO/UKST SuperCOSMOS H-alpha Survey (SHS, Parker et al. 2005). An international workshop in honour of Agnès Acker was held at the ATNF/AAO from February 16–18 to celebrate the Macquarie/ AAO/Strasbourg H-alpha (MASH) planetary nebulae catalogues (Parker et al. 2006; Miszalski et al. 2008) that have discovered over 1200 new planetary nebulae solely from careful examination of the SHS.

Epping Lab
Telephone +61 2 9372 4800 Fax +61 2 9372 4880 email < *user*@aao.gov.au >

AAT/Schmidt
Telephone +61 2 6842 6291 Fax +61 2 6884 2298 email < *user*@aaocbn.aao.gov.au >

URL < http://www.aao.gov.au >